\documentclass{article}
\usepackage{amsmath,amssymb,bbm,amsthm}
\usepackage{fullpage}
\usepackage{thm-restate,color,xcolor,xspace}
\usepackage{hyperref,cleveref}
\usepackage{algorithm,algorithmicx}
\usepackage[noend]{algpseudocode}
\usepackage{mathtools}
\usepackage{tikz}

\newcommand{\f}{\frac}
\newcommand{\cd}{\cdot}
\newcommand{\bn}{\binom}
\newcommand{\sr}{\sqrt}
\newcommand{\cds}{\cdots}
\newcommand{\lds}{\ldots}
\newcommand{\vds}{\vdots}
\newcommand{\dds}{\ddots}
\newcommand{\pge}{\succeq}
\newcommand{\ple}{\preceq}
\newcommand{\bs}{\setminus}
\newcommand{\s}{\subseteq}

\newcommand{\BE}{\begin{enumerate}}
\newcommand{\EE}{\end{enumerate}}
\newcommand{\im}{\item}
\newcommand{\BI}{\begin{itemize}}
\newcommand{\EI}{\end{itemize}}

\newcommand{\Sum}{\displaystyle\sum\limits}
\newcommand{\Prod}{\displaystyle\prod\limits}
\newcommand{\Int}{\displaystyle\int\limits}
\newcommand{\Lim}{\displaystyle\lim\limits}
\newcommand{\Max}{\displaystyle\max\limits}
\newcommand{\Min}{\displaystyle\min\limits}

\newcommand{\logn}{\log n}

\newcommand{\dx}{\frac d{dx}}
\newcommand{\dy}{\frac d{dy}}
\newcommand{\dz}{\frac d{dz}}
\newcommand{\dt}{\frac d{dt}}

\newcommand{\inv}{^{-1}}

\newcommand{\R}{\mathbb R}
\newcommand{\Z}{\mathbb Z}
\newcommand{\F}{\mathbb F}
\newcommand{\C}{\mathbb C}
\newcommand{\N}{\mathbb N}
\newcommand{\Q}{\mathbb Q}

\newcommand{\eps}{\epsilon}
\newcommand{\e}{\epsilon}
\newcommand{\de}{\delta}
\newcommand{\De}{\Delta}
\newcommand{\la}{\lambda}
\newcommand{\g}{\gamma}
\newcommand{\G}{\Gamma}
\newcommand{\pt}{\partial}
\newcommand{\al}{\alpha}
\newcommand{\be}{\beta}
\newcommand{\om}{\omega}
\newcommand{\Om}{\Omega}
\newcommand{\el}{\ell}
\renewcommand{\th}{\theta}
\newcommand{\Th}{\Theta}
\newcommand{\m}{\mathcal}

\newcommand{\Ra}{\Rightarrow}

\newcommand{\lf}{\lfloor}
\newcommand{\rf}{\rfloor}
\newcommand{\lc}{\lceil}
\newcommand{\rc}{\rceil}

\newcommand{\E}{\mathbb E}
\newcommand{\Var}{\text{Var}}
\newcommand{\Cov}{\text{Cov}}
\newcommand{\1}{\mathbbm 1}
\newcommand{\poly}{\text{poly}}
\newcommand{\polylog}{\text{polylog}}
\newcommand{\norm}[1]{\left\lVert#1\right\rVert}

\newcommand{\rank}{\text{rank}}
\newcommand{\spn}{\text{span}}
\newcommand{\Tr}{\text{Tr}}

\newcommand{\lp}{\left(}
\newcommand{\rp}{\right)}
\newcommand{\lb}{\left[}
\newcommand{\rb}{\right]}
\newcommand{\lmt}{\left[\begin{matrix}}
\newcommand{\rmt}{\end{matrix}\right]}

\newtheorem{theorem}{Theorem}[section]
\newtheorem{lemma}[theorem]{Lemma}
\newtheorem{definition}[theorem]{Definition}
\newtheorem{corollary}[theorem]{Corollary}
\newtheorem{observation}[theorem]{Observation}
\newtheorem{claim}[theorem]{Claim}
\newtheorem{assumption}[theorem]{Assumption}
\newtheorem{fact}[theorem]{Fact}
\newcommand{\BT}{\begin{theorem}}
\newcommand{\ET}{\end{theorem}}
\newcommand{\BL}{\begin{lemma}}
\newcommand{\EL}{\end{lemma}}
\newcommand{\BD}{\begin{definition}}
\newcommand{\ED}{\end{definition}}
\newcommand{\BC}{\begin{corollary}}
\newcommand{\EC}{\end{corollary}}
\newcommand{\BO}{\begin{observation}}
\newcommand{\EO}{\end{observation}}
\newcommand{\BCL}{\begin{claim}}
\newcommand{\ECL}{\end{claim}}
\newcommand{\BP}{\begin{proof}}
\newcommand{\EP}{\end{proof}}
\newcommand{\BPS}{\begin{proof}[Proof (Sketch)]}
\newcommand{\EPS}{\end{proof}}
\newcommand{\BA}{\begin{assumption}}
\newcommand{\EA}{\end{assumption}}
\newcommand{\BF}{\begin{fact}}
\newcommand{\EF}{\end{fact}}
\Crefname{observation}{Observation}{Observations}
\Crefname{assumption}{Assumption}{Assumptions}
\Crefname{claim}{Claim}{Claims}
\Crefname{subclaim}{Subclaim}{Subclaims}
\Crefname{assumption}{Assumption}{Assumptions}
\Crefname{fact}{Fact}{Facts}

\newtheorem{alg}[theorem]{Algorithm}
\newcommand{\BAL}{\begin{alg}}
\newcommand{\EAL}{\end{alg}}
\Crefname{alg}{Algorithm}{Algorithms}

\newcommand{\alert}{\textcolor{red}}
\newcommand{\para}{\paragraph}
\newcommand{\defn}{\textbf}

\newcommand{\tO}{\tilde{O}}
\newcommand{\ol}{\overline}
\newcommand{\vol}{\textup{\textbf{vol}}}

\newcommand{\kcut}{\ensuremath{k\textsc{-Cut}}\xspace}
\newcommand{\kpath}{\ensuremath{k\textsc{-Path}}\xspace}

\makeatletter
\newcounter{algocounter}
\@ifpackageloaded{hyperref}%
  {\newcommand{\mylabel}[2]
    {\refstepcounter{algocounter}\protected@write\@auxout{}{\string\newlabel{#1}{{\textcolor{black}{\textup{#2}}}{\thepage}%
      {\@currentlabelname}{\@currentHref}{}}}}}%
\makeatother

\newcommand{\mincut}{\mathsf{MinCut}}
\newcommand{\minanccut}{\mathsf{MinAncCut}}
\newcommand{\minelts}{\mathsf{MinElts}}
\newcommand{\opt}{\mathsf{Opt}}
\newcommand{\Time}{\mathsf{Time}}
\newcommand{\Ttree}{\textup{T-tree}\xspace}
\newcommand{\depth}{\textup{depth}\xspace}
\newcommand{\orig}{\textup{orig}\xspace}
\newcommand{\col}{\textsf{color}\xspace}
\newcommand{\st}{\textsf{\textup{State}}\xspace}
\newcommand{\subroot}{\textsf{\textup{subroot}}\xspace}
\newcommand{\oa}{\overrightarrow}
\newcommand{\ra}{\textsf{\textup{rank}}\xspace}

\newcommand{\red}{{\textsf{\textup{red}}}\xspace}
\newcommand{\green}{{\textsf{\textup{green}}}\xspace}

\newtheorem{cond}[theorem]{Condition}
\newcommand{\BCO}{\begin{cond}}
\newcommand{\ECO}{\end{cond}}
\Crefname{cond}{Condition}{Conditions}

\begin{document}

\title{Faster Minimum $k$-cut of a Simple Graph}
\author{Jason Li \\ Carnegie Mellon University \\ jmli@cs.cmu.edu}
\date{\today}
\maketitle

\abstract{
We consider the (exact, minimum) \kcut problem: given a graph and an integer $k$, delete a minimum-weight set of edges so that the remaining graph has at least $k$ connected components. This problem is a natural generalization of the global minimum cut problem, where the goal is to break the graph into $k=2$ pieces.

Our main result is a (combinatorial) \kcut algorithm on simple graphs that runs in $n^{(1+o(1))k}$ time for any constant $k$, improving upon the previously best $n^{(2\om/3+o(1))k}$ time algorithm of Gupta et al.~[FOCS'18] and the previously best $n^{(1.981+o(1))k}$ time combinatorial algorithm of Gupta et al.~[STOC'19]. For combinatorial algorithms, this algorithm is optimal up to $o(1)$ factors assuming recent hardness conjectures: we show by a straightforward reduction that \kcut on even a simple graph is as hard as $(k-1)$-clique, establishing a lower bound of $n^{(1-o(1))k}$ for \kcut. This settles, up to lower-order factors, the complexity of \kcut on a simple graph for combinatorial algorithms.
}

\section{Introduction}
We consider the (exact, minimum) \kcut problem: given a graph and an integer $k$, delete a minimum-weight set of edges so that the remaining graph has at least $k$ connected components. This problem is a natural generalization of the global minimum cut problem, where the goal is to break the graph into $k=2$ pieces. This problem has been actively studied in theory of both exact and approximation algorithms, where each result brought new insights and tools on graph cut algorithms.

Goldschmidt and Hochbaum gave the first polynomial-time algorithm for
fixed $k$, with $O(n^{(1/2 - o(1))k^2})$ runtime~\cite{GH94}. Since
then, the exact exponent in terms of $k$ has been actively studied.  The
textbook minimum cut algorithm of Karger and Stein~\cite{KS96}, based on
random edge contractions, can be adapted to solve \kcut in
$\tO(n^{2(k-1)})$ (randomized) time.
This bound was improved recently for the first time by an algorithm of Gupta et al.~\cite{GLL19}, which runs in $n^{(1.981+o(1))k}$ (randomized) time.
The deterministic algorithms side
has seen a series of improvements since
then~\cite{KYN06,Thorup08,chekuri2018lp}. The fastest algorithm for
general edge weights is due to Chekuri et al.~\cite{chekuri2018lp}. It
runs in $O(mn^{2k-3})$ time and is based on a deterministic tree packing
result of Thorup~\cite{Thorup08}.
Lastly, if the edge weights of the input graph are integers bounded by $n^{O(1)}$ (in particular, exponent independent of $k$), then \kcut can be solved in $n^{(2\om/3+o(1))k}$ deterministic time~\cite{GLL18b}, where $\om<2.373$ is the matrix multiplication constant~\cite{le2014powers,williams2012multiplying}.

Lower bounds for the \kcut problem have also been studied actively in the past decade. \kcut on real-weighted graphs is at least as hard as minimum weighted $(k-1)$-clique~\cite{GLL18b}, the latter of which is conjectured to require $n^{(1-o(1))k}$ time for any constant $k$~\cite{williams2010subcubic}. For \kcut on unweighted graphs, the lower bound is weakened to $n^{(\om/3-o(1))k}$, again from a reduction to $(k-1)$-clique, where $\om<2.3727$ is the matrix multiplication constant~\cite{GLL18b}. However, for ``combinatorial'' algorithms as described in~\cite{abboud2015if,williams2010subcubic}, this lower bound is again $n^{(1-o(1))k}$ even for unweighted graphs, under the stronger hardness conjecture of $k$-clique for combinatorial algorithms~\cite{abboud2015if,williams2010subcubic}.

In this paper, we consider the \kcut on \emph{simple} graphs: graphs that are unweighted and have no parallel edges. Our main result is a (combinatorial) \kcut algorithm on simple graphs that runs in $n^{(1+o(1))k}$ time for any constant $k$, improving upon the previously best $n^{(2\om/3+o(1))k}$ time algorithm \cite{GLL18b} and the previously best $n^{(1.981+o(1))k}$ time combinatorial algorithm~\cite{GLL19}. For combinatorial algorithms, this algorithm is optimal up to $o(1)$ factors assuming recent hardness conjectures: we show by a straightforward reduction that \kcut on even a simple graph is as hard as $(k-1)$-clique, establishing a lower bound of $n^{(1-o(1))k}$ for \kcut. This settles, up to lower-order factors, the complexity of \kcut on a simple graph for combinatorial algorithms. We remark that this is the first setting for \kcut, except the restricted $k=2$ case, where the running time has been determined up to $o(1)$ factors.

\BT[Main Result]\label{thm:main}
For any parameter $k$, there is a (combinatorial, randomized) algorithm that computes the \kcut of a simple graph in $k^{O(k)}n^{(1+o(1))k}$ time.
\ET
\BT[Lower Bound]\label{thm:lower}
Suppose we assume the conjecture that every combinatorial algorithm for $k$-clique requires $n^{(1-o(1))k}$ time for any constant $k$. Then, for any constant $k$, every combinatorial algorithm for \kcut of a simple graph also requires $n^{(1-o(1))k}$ time.
\ET


\subsection{Our Techniques}\label{thm:tec}

Our \kcut algorithm incorporates algorithmic techniques from a wide array of areas, from graph sparsification to fixed-parameter tractability to tree algorithms.  

\para{Graph Sparsification}
Our first algorithmic ingredient is the Kawarabayashi-Thorup (KT) sparsification algorithm, which originated from the breakthrough paper of Kawarabayashi and Thorup on the deterministic minimum cut problem~\cite{KT}. At a high level, given any simple graph $G$ with minimum cut $\la$, the algorithm contracts $G$ into a multi-graph of $\tO(m/\la)$ \footnote{Throughout the paper, we use the standard $\tO(\cd)$ notation to hide polylogarithmic factors in the running time.} edges so that any minimum cut of $G$ that has at least two vertices on each side gets ``preserved'' in the contraction. That is, we never contract an edge in any such minimum cut. Kawarabayashi and Thorup used their contraction procedure to provide the first $\tO(m)$-time deterministic algorithm for minimum cut of a simple graph. They first applied the contraction to $G$, obtaining a multi-graph $\ol G$ on $\ol{m}=\tO(m/\la)$ edges, and then ran the $\tO(\ol m\la)$-time minimum cut algorithm of Gabow on $\ol G$, which works for multi-graphs. This covers the case when the minimum cut of $G$ has at least two vertices on each side; the other case, where the minimum cut consists of a singleton vertex on one side, is trivial. In the KT sparsification, we can also ensure that $\ol G$ has $\tO(n/\la)$ vertices, which is the property that we focus on.

One of our technical contributions is extending the KT sparsification algorithm to work for \kcut. Briefly, we show that modulo a regularity condition, we can contract the input graph $G$ into a multi-graph $\ol G$ on $\ol n=\tO(n/\la_k)$ vertices, where $\la_k$ is the minimum $k$-cut of $G$, to preserve all minimum $k$-cuts where each side has at least two vertices. (For the case when one component of the minimum $k$-cut is a singleton vertex, we handle it separately with a simple branching procedure: try each vertex as a singleton component and recursively solve $(k-1)$-cut.) In the spirit of Kawarabayashi and Thorup, we then solve \kcut on the contracted graph in $\ol n^{(1+o(1))k}\la_k^k$ time. Since $\ol n=\tO(n/\la_k)$, this running time becomes $n^{(1+o(1))k}$, as needed.

\para{Tree Packing}
To solve \kcut in $n^{(1+o(1))k}\la_k^k$ time on multi-graphs,\footnote{Here, we assume the contracted graph is our input graph now, hence $n$ instead of $\ol n$.} we begin with a tree packing result of Thorup~\cite{Thorup08}, which says that we can compute a small collection of trees so that for one tree $T$, at most $2k-2$ edges of $T$ have endpoints in different components of the minimum $k$-cut. we apply a reduction in \cite{GLL18b} which, at a multiplicative cost of $O(n^k)$ in the running time, produces a tree $T$ such that exactly $k-1$ edges of $T$ have endpoints in different components of the minimum $k$-cut. (Note that $k-1$ here is the smallest possible.) In other words, if we remove these edges from $T$, then the connected components in the remaining forest are exactly the components of the minimum $k$-cut. 
\para{Color Coding}
Following the last paragraph, our problem thus reduces to this: given a graph $G=(V,E)$ and a tree $T$ on the vertices $V$, remove some $(k-1)$ edges $F$ of $T$ to minimize the $k$-cut (in $G$) formed by the $k$ connected components in $T-F$.
 Our main technical contribution is providing such an algorithm that runs in $n^{o(k)}\la^k$ time.

One major ingredient in our algorithm is the technique of \emph{color coding} due to Alon et al.~\cite{alon1995color}, who first used it for the \kpath problem in the fixed-parameter setting.  To illustrate our approach, let us assume (\emph{with} loss of generality, for sake of exposition) that the tree $T$ is a ``spider'': it can be decomposed into an edge-disjoint set of paths sharing a common endpoint $r$; see the black edges in Figure~\ref{f1} for an example. Call each of the edge-disjoint paths from $r$ be a \emph{branch}. Let us further assume (again, \emph{with} loss of generality) that the optimal set $F^*$ of $(k-1)$ edges consists of one edge from each of $(k-1)$ distinct branches $B^*_1,\lds,B^*_{k-1}$. Let $S^*_1,\lds,S^*_k\s V$ be the components of $T-F^*$ with $r\in S^*_k$, and let $OPT\s E$ be the minimum $k$-cut (in $G$) with components $S^*_1,\lds,S^*_k$.

Our first observation is that if there were no edges between $S^*_i$ and $S^*_j$ for every $1\le i<j\le k-1$, then the problem becomes easy, because the following (polynomial-time) algorithm works:
\BE
\im For each branch $B$, pick the edge $e$ in the branch to remove so that the two connected components $C_1,C_2\s V$ of $T-e$ minimize $|E[C_1,C_2]|$ (the number of edges in $G$ between $C_1$ and $C_2$). Let $f(B)$ be the minimum value $|E[C_1,C_2]|$ for branch $B$.
\im Select the $(k-1)$ branches $B_1,\lds,B_{k-1}$ with the smallest values of $f(B_i)$. The $(k-1)$ edges to remove are the edges selected in each of these branches, and the total $k$-cut value is $\sum_{i=1}^{k-1}f(B_i)$.
\EE
In other words, the algorithm processes each branch independently and selects the best $(k-1)$ branches.

In general, if there is an edge in the minimum $k$-cut between $S_1$ and $S_2$ (say), then this edge may contribute to both $f(B_1)$ and $f(B_2)$, in which case it is double-counted. So we always have $\sum_{i=1}^{k-1}f(B_i)\ge|OPT|$, and strict inequality is possible. But if every edge in $OPT$ connects $f(B_k)$ to some $f(B_i)$ ($i\le k-1$), then every edge is counted at most once, so $\sum_{i=1}^{k-1}f(B_i)=OPT$.

 This algorithm works in this special setting because no ``double-counting'' occurs: every edge in $OPT$ is counted exactly once in $\sum_{i=1}^{k-1}f(B_i)$. What if we consider the opposite case scenario, where \emph{many} edges are double-counted? In particular, suppose that there is at least one edge between every two branches $B^*_i,B^*_j$ with $i,j\le k-1$. Let $E_2\s OPT$ be the edges in $OPT$ in between two such branches.

These edges may fool the simple algorithm described above, but they serve a different useful purpose. Namely, the edges $E_2$ \emph{connect} the branches $B^*_1,\lds,B^*_k$ together, and that is a property we will exploit as follows: Let us randomly color each edge of $E$ either red or green, hoping for the following two properties:
\BE
\im For each $i<j\le k-1$, there is at least one edge in $E_2$ colored green.
\im All edges in $E\setminus E_2$ incident to vertices in $B^*_1,\lds,B^*_k$ are colored red.
\EE
The properties we require are a bit more specific, but the two conditions above suffice for illustration. If both properties as satisfied, then if we consider the connected components of branches, where two branches are pairwise connected if they share a green edge, then $B^*_1,\lds,B^*_k$ exactly mark out a single component. And if the algorithm iterates over all such components of branches (and processes each one independently, say), then it will come across exactly $\{B^*_1,\lds,B^*_k\}$ at some point. Thus, in some sense, we may \emph{assume} that the algorithm knows $B^*_1,\lds,B^*_k$.

Even with this knowledge, the issue of double-counting still remains. To handle it, we pinpoint down a small set of edges $E'\s E_2$ such that if all edges in $E'$ are green, then the algorithm learns enough information about the double-counted edges to provide the correct answer. The size of $E'$ needs to be small, because we color each edge green with small probability, and yet require that all of $E'$ is colored green with sufficiently large probability.

\para{Tree Algorithms}
The algorithm above only handles the case when $T$ is a ``spider''. What about the general case? Our first idea is to apply \emph{heavy-light decomposition} on the (rooted) tree, breaking it into a disjoint union of branches, with the property that every path from leaf to root intersects the edges of $O(\logn)$ branches. This $O(\log n)$ factor will be paid in the running time as $O(\log n)^k$, which (perhaps surprisingly) can be bounded by $\max\{k^{O(k)},n\}$; morally, this means that the $O(\logn)$ factor is negligible. The benefit of the HLD is that we once again have a disjoint union of branches. Although these branches may not have a common endpoint, they \emph{almost} do, in the sense that every path from leaf to root now intersects $O(\logn)$ branches instead of $1$, which is still small.

Lastly, what about the case when multiple edges are cut from the same branch? To handle this situation, we apply dynamic programming on a tree in a manner similar to~\cite{GLL18b}. The key observation is that if we remove an edge $e$ from $T$, then the two components of $T_e$ become independent subproblems, thus making the situation amenable to dynamic programming.

\subsection{Related Work}
The \kcut problem has been studied extensively in the approximate and fixed-parameter settings as well.
\paragraph{Approximation algorithms.} The first approximation algorithm \kcut was a
$2(1-1/k)$-approximation of Saran and Vazirani~\cite{SV95}.  Later, Naor
and Rabani~\cite{NR01}, and also Ravi and Sinha~\cite{RS02} gave
$2$-approximation algorithms using tree packing and network strength
respectively.  Xiao et al.~\cite{XCY11} extended Kapoor~\cite{Kapoor96}
and Zhao et al.~\cite{ZNI01} and generalized Saran and Vazirani to give
an $(2 - h/k)$-approximation in time $n^{O(h)}$. On the hardness front,
Manurangsi~\cite{Manurangsi17} showed that for any $\eps > 0$, it is
NP-hard to achieve a $(2 - \eps)$-approximation algorithm in time
$\poly(n,k)$ assuming the Small Set Expansion Hypothesis.

Recently~\cite{GLL18b}, Gupta et al.\ gave a $1.81$-approximation for
\kcut in FPT time $f(k) \poly(n)$ and a $(1+\e)$-approximation in $f(k)n^{(1+o(1))k}$ time. These algorithms do not contradict Manurangsi's
work, since $k$ is polynomial in $n$ for his hard instances. 

\paragraph{FPT algorithms.}
The \kcut problem was shown to be $W[1]$-hard when parameterized by $k$ by Downey et al.~\cite{DEFPR03}. 
Kawarabayashi and Thorup give the first $f(\textsf{Opt})
\cdot n^{2}$-time algorithm~\cite{KT11} for unweighted graphs. Chitnis
et al.~\cite{Chitnis} used a randomized color-coding idea to give a
better runtime, and to extend the algorithm to weighted graphs. Here,
the FPT algorithm is parameterized by the cardinality of edges in the
optimal \kcut, not by the number of parts $k$. For more details on FPT
algorithms and approximations, see the book~\cite{FPT-book}, and the 
survey~\cite{Marx07}.

\section{Preliminaries}

All graphs in this paper will be unweighted, undirected multigraphs without self-loops. We denote by $n$ and $m$ the number of vertices and edges in the input graph, respectively. A graph is \emph{simple} if for every two vertices in the graph, there is at most one edge between them. For a graph $G$, let $V(G)$ and $E(G)$ be its vertex set and edge set, respectively. For vertex-disjoint sets $S_1,\lds,S_\el$, denote by $E[S_1,\lds,S_\el]$ the set of edges whose endpoints lie in distinct sets $S_i,S_j$ ($i\ne j$). When there are multiple graphs in our context, we use $E_G[S_1,\lds,S_\el]$ instead to indicate that the graph in question is $G$; we use similar notation for other graph functions. For a vertex set $S\s V(G)$ of a graph $G$, denote $\pt S=E[S,V\setminus S]$ as the set of edges with exactly one endpoint in $S$. For disjoint vertex sets $S_1,\lds,S_\el\s V(G)$, denote $\pt[S_1,\lds,S_\el]=E[S_1,\lds,S_\el,V\setminus\bigcup_{i=1}^\el S_i]$  as the set of edges with at least one endpoint in some $S_i$, but not both endpoints in the same $S_i$; note that $\pt S=\pt[S]$ for any subset $S\s V(G)$. The \emph{degree} of a vertex $v\in V(G)$ is the number of edges incident to it, which equals $|\pt(\{v\})|$.

For a graph $G=(V,E)$, the minimum $k$-cut will either be denoted as the subsets $S^*_1,\lds,S^*_k\s V$ that comprise the components of the $k$-cut, or as the edge set $OPT=E[S^*_1,\lds,S^*_k]$, the set of edges in the $k$-cut. We now define the concept of a \emph{nontrivial} minimum $k$-cut:

\BD[Nontrivial minimum $k$-cut]
A minimum $k$-cut $\{S^*_1,\lds,S^*_k\}$ is \emph{nontrivial} if none of the sets $|S^*_i|$ ($i\in[k]$) have size $1$.
\ED

Since we work with multigraphs throughout the paper, every edge has a unique identifier. Whenever we declare a variable $e$ as an edge, we mean its identifier. In particular, if two edges $e,e'$ both have endpoints $u,v$, then it is possible that $e\ne e'$. We identify each edge by its identifier, rather than its two endpoints $(u,v)$. That is, every variable $e$ designated to an edge is set to the edge's identifier, rather than the tuple $(u,v)$ or the set $\{u,v\}$. We may still say ``an edge $(u,v)\in E(G)$'', by which we mean an arbitrary edge with endpoints $u,v$ in $G$ (and we do not care about its identifier). Likewise, ``for every edge $(u,v)\in E(G)$'' means every edge with endpoints $u,v$ in $G$.

 Whenever we \emph{contract} two vertices $u,v\in V(G)$ in a graph $G$, all edges that survive keep their identifiers. More formally, contraction produces the following graph $G'$:
\BE
\im We have $V(G')=V(G)\setminus\{u,v\}\cup\{x\}$ for some new vertex $x$.
\im For every edge $e$ with endpoints $u',v'$ distinct from $u$ and $v$, add edge $e$ with the same endpoints $u',v'$ in $G'$.
\im For every edge $e$ with endpoints $u,v'$ for $v'\ne v$, add edge $e$ with endpoints $x,v'$ in $G'$.
\im For every edge $e$ with endpoints $u',v$ for $u'\ne u$, add edge $e$ with endpoints $u',x$ in $G'$.
\im For every edge $e$ with endpoints $u,v$ in $G$, do not add it to $G'$.
\EE
In particular, the edge identifiers in $G$ and $G'$ match. One benefit to this formulation is that if $OPT\s E$ is a minimum $k$-cut with components $S^*_1,\lds,S^*_k$ and we contract two vertices $u,v\in S^*_i$ for some $i$, then the same set $OPT$ is still a minimum $k$-cut in the contracted graph.

For a positive integer $\el$, we denote by $[\el]$ the set of integers from $1$ to $\el$ (inclusive), $\{1,2,\lds,\el\}$.

\para{Tree Terminology (\Cref{sec:tree})}
The terminology in this paragraph are specific to \Cref{sec:tree}.

For a rooted tree $T$, let $T(v)\s T$ denote the subtree of $T$ rooted at $v\in V(T)$. For any set $S \s V(T)$, define $T(S) := \bigcup_{v \in S} T(v)$, the union of all (vertices and edges of) trees $T(v)$ over all $v\in S$.

For an edge $e=(u,v)\in V(T)$ where $v$ is the child of $u$, we say that $v$ is the \emph{child vertex} of edge $e$ and $e$ is the \emph{parent edge} of $v$.


Given a rooted tree $T$, the \emph{depth} $\depth(v)$ of vertex $v\in V(T)$ is the (unweighted) distance of $v$ to the root of $T$. Every time we use $\depth(v)$, the tree $T$ will be implicit.

A \emph{branch} of $T$ is a path in $T$ that travels ``downwards'' the rooted tree. More formally, it is a path whose vertices have distinct depths. The vertex with the minimum depth is the \emph{root} of the branch. A \emph{maximal} branch is a branch from the root to a leaf in the tree. When we say a \emph{branch from $u$ to $v$}, we mean the path from $u$ and $v$.

For a rooted tree $T$ and $u,v\in V(T)$, we say that $u$ \emph{precedes} $v$ if $v\in T(u)$. We say that two vertices $u,v\in V(T)$ are \emph{incomparable} if $u$ does not precede $v$ and $v$ does not precede $u$. Note that $u$ and $v$ are incomparable iff  they do not lie on a common branch.

Given a set $S\s V(T)$, denote by $S_{\downarrow}$ the \emph{minimal elements of $S$}, the minimal set $S'\s S$ such that every vertex in $S$ is preceded by some vertex in $S'$ (that is, $S\s T(S')$).

\para{Tree Packing.}
Our algorithm will use the concept of \emph{tree packing} which, at a high level, reduces the \kcut to a problem of finding the best way to remove edges in a tree. Tree packings for the \kcut problem were first introduced by Thorup~\cite{Thorup08}, who used them to obtain a deterministic \kcut algorithm in time $O(mn^{2k-2})$.

\BD[\Ttree, Definition~2.1 of \cite{GLL18b}]\label{def:treepacking}
A tree $T$ of $G$ is a $\ell$-\Ttree if it crosses some
  minimum $k$-cut at most $\ell$ times; i.e., $E_T(S^*_1,\lds,S^*_k) \leq
  \ell$. If $\ell = k-1$, the minimum value possible, then we call it
  a tight \Ttree.

\ED

\BT[Thorup \cite{Thorup08}, rephrased in Corollary~2.3 of \cite{GLL18b}]\label{2.3}
We can find a collection $\m T$ of $\tO(k^3m)$ trees such that there exists a $(2k-2)$-\Ttree in $\m T$.
\ET

\BT[Lemma~2.4~of~\cite{GLL18b}]\label{2.4}
  There is an algorithm that takes as input a tree $T$ such that
  $|E_T(S^*_1,\lds,S^*_k)|\le2k-2$, and produces a collection of
  $k^{O(k)}n^{k-1}\log n$ trees, such that one of the new trees $T'$
  satisfies $|E_{T'}(S^*_1,\lds,S^*_k)|=k-1$ w.h.p. The
  algorithm runs in time $k^{O(k)}n^{k-1}m\log n$.
\ET

Combining \Cref{2.3} and \Cref{2.4}, we obtain the following:
\BC\label{treepacking}
We can find a collection $\m T$ of $k^{O(k)}n^{k+O(1)}$ trees such that there exists a tight \Ttree in $\m T$.
\EC
For the rest of the paper, only \Cref{def:treepacking} and \Cref{treepacking} will be used.

\section{Algorithm Outline}

In this section, we outline our main algorithm, assuming our two main technical results below. The former is proved in \Cref{sec:tree} and the latter in \Cref{sec:KT}.

\begin{restatable}{theorem}{tree}\label{thm:tree}
Let $G$ be an unweighted multigraph, let $T$ be a tight \Ttree of $G$, and let $s$ be a parameter. There is an algorithm \textup{TreeCut}$(G,T,\la)$ with the following guarantee: if the minimum $k$-cut in $G$ has size~$\le \la$, then \textup{TreeCut} outputs a minimum $k$-cut of $G$. The running time of \textup{TreeCut} is $k^{O(k)}\la^kn^{o(k)}$.
\end{restatable}

\begin{restatable}{theorem}{KT}\label{thm:KT}
Let $G$ be a simple graph with minimum degree $\de > \om(\max\{\al\logn,\al k\})$, and let $\al\ge1$ be a parameter. Then, we can contract $G$ into a (multi-)graph $\ol G$ such that:
  \BE
  \im Suppose the minimum $k$-cut has size $\le\al\de$ in $G$. Then, every  nontrivial minimum $k$-cut is preserved in $\ol G$. That is, no edge of such a cut is contracted in $\ol G$.
  \im $\ol G$ has $\tO(\al m/\de)$ edges and $\tO(\al m/\de^2)$ vertices.
  \EE
\end{restatable}

We will also use the following sparsification routine due to Nagamochi and Ibaraki below, for which we provide a quick proof for self-containment:

\BT[Nagamochi-Ibaraki~\cite{NI92}]\label{thm:NI}
Given a simple graph $G$ and parameter $\la$, there is a polynomial-time algorithm \textup{NISparsify}$(G,\la)$ that computes a subgraph $H$ with at most $\la n  $ edges such that all $k$-cuts of size $\le\la$ are preserved. More formally, all sets $S$ with $|\pt_GS|\le\la$ satisfy $|\pt_GS|=|\pt_HS|$.
\ET
\BP
For $i=1,2,\lds,\la$, let $F_i$ be a maximal forest in $G\setminus\bigcup_{j<i}F_j$. Set $H:=\bigcup_iF_i$, which can easily be computed in polynomial time. (We note that Nagamochi and Ibaraki~\cite{NI92} present a way to compute $H$ in \emph{linear} time, although we do not need this.) For any edge $(u,v)$ in $G-H=G\setminus \bigcup_iF_i$, there must be a $(u,v)$ path in each $F_i$, otherwise we would have added edge $(u,v)$ to $F_i$. These $\la$ paths, along with edge $(u,v)$, imply that every cut that separates $u$ and $v$ has size~$\ge\la+1$.  Therefore, $u$ and $v$ must lie in the same component of any $k$-cut of size~$\le\la$, so removing edge $(u,v)$ cannot affect any such $k$-cut.
\EP

Let us now describe our algorithm in pseudocode:

\newcommand{\UPDATE}[1]{If $\m S^#1$ is a better $k$-cut than $\m S^*$, then set $\m S^*\gets\m S^#1$}
\begin{algorithm}[H]
\mylabel{K}{\texttt{MinKCut}}
\caption{\ref{K}($G=(V,E),k$)}
\begin{algorithmic}[1]
\If {$k=1$} \Comment{Base case $k=1$}
  \State\Return $\{V\}$
\EndIf
\State $\m S^* \gets \{S^0_1,\lds,S^0_k\}$, an arbitrary initial $k$-cut \Comment{$k$-cuts will be represented as partitions of $V$ of size $k$}
\For {each $v\in V$}
  \State $\m S^1\gets \{v\} \cup \ref K(G-v, k-1$)\label{line:m3}\Comment{Recursively call minimum $(k-1)$-cut}
  \State \UPDATE1
\EndFor
\If {$\de > \om(\max\{k^2\logn,k^3\})$} \label{line:m10} \Comment{Assumption of \Cref{thm:KT}. $\de$ is the minimum degree of $G$}
  \State $H\gets\textup{NISparsify}(G,k^2\de)$ \label{line:NI} \Comment{Nagamochi-Ibaraki sparsification: see \Cref{thm:NI}}
  \State $G\gets \ref{KT}(H, k^2)$ \Comment{Replace $G$ with the KT-sparsification of $G$ (\Cref{thm:KT})}
\EndIf 
\State $\m T \gets \text{TreePacking}(G,k)$\Comment{\Cref{treepacking}} \label{line:m12}
\For {each tree $T\in\m T$}
\State $\m S^T\gets \text{TreeCut}(G,T,k^2\de)$\label{line:treecut}\Comment{\Cref{thm:tree}}
\State \UPDATE T
\EndFor
\State\Return $\m S^*$
\end{algorithmic}\label{alg:KT}
\end{algorithm}

\subsection{Correctness}

We first state an easy claim from~\cite{GLL18b}. We then use it to bound the size of a nontrivial minimum $k$-cut by the minimum degree~$\de$ of the graph.
\BCL[Claim~2.8 of \cite{GLL18b}]\label{2.8}
Given a set of $k+1$ components $S_1,\lds,S_{k+1}$ that partition $V$, we have
\[ |OPT| \le \lp1-\bn{k+1}2\inv\rp|E[S_1,\lds, S_{k+1}]| .\]
\ECL

\BL\label{lem:nont}
Suppose there exists a nontrivial minimum $k$-cut in the graph. Then, we have $|OPT|\le k^2\de$.
\EL
\BP
Fix a nontrivial minimum $k$-cut $S^*_1,\lds,S^*_k$. Applying \Cref{2.8} on $S_i:=S^*_1\setminus\{v\}$ for $i\in[k]$ and $S_{k+1}=\{v\}$, we get
\[ |OPT| < \lp1-\f1{k^2}\rp |E[S_1,\lds,S_{k+1}]|\le \lp1-\f1{k^2}\rp (|E[S^*_1,\lds,S^*_k]|+\de)= \lp1-\f1{k^2}\rp (|OPT|+\de) ,\] 
so $|OPT| \le k^2(1-k^2)\de \le k^2\de$, as needed.
\EP

\BL[Correctness]
$\ref K(G,k)$ outputs a minimum $k$-cut of $G$.
\EL
\BP
If there is a nontrivial minimum $k$-cut $v$, then consider the iteration of line~\ref{line:m3} on the vertex $v$. By induction on $k$, we may assume that \ref K$(\cd,k-1)$ outputs a minimum $k$-cut. Therefore, $\m S^1$ will be an optimal $k$-cut on this iteration of line~\ref{line:m3}.

Otherwise, suppose there is no nontrivial minimum $k$-cut, so by \Cref{lem:nont}, $|OPT|\le k^2\de$. First, suppose that line~\ref{line:m10} holds. Then, by \Cref{thm:NI} with $s:=k^2\de$, the graph $H$ in line~\ref{line:NI} has the same minimum $k$-cuts as $G$ and has at most $k^2\de n$ edges. By \Cref{thm:KT} with $\al:=k^2$, every (nontrivial) minimum $k$-cut in $H$ also exists in $\ref{KT}(H, k^2)$, which we set as our new $G$. By the correctness of TreePacking$(G,k)$ (\Cref{treepacking}) and TreeCut$(G,T)$ (\Cref{thm:tree}), the algorithm computes an optimal $k$-cut. Otherwise, if  line~\ref{line:m10} does not hold, then our situation is even easier, since $G$ does not change.
\EP

\subsection{Running time}
\BF\label{lognk}
$(\log n)^{O(k)} \le \max\{k^{O(k)},n\}$.
\EF
\BP
If $k<\f\logn{\log\logn}$, then
$(\log n)^k \leq  \smash{(\log
n)^{\f\logn{\log\logn}}} =  n$. Else $\log n \leq O(k
\log k)$, and hence $(\log n)^k \le O(k\log k)^k\le k^{O(k)}$.
\EP

\BL
$\ref K(G,k)$ runs in $k^{O(k)}n^{(1+o(1))k}$ time.
\EL
\BP
First, we bound the running time outside the recursive calls in line~\ref{line:m3}. Suppose first that  line~\ref{line:m10} does not hold. Then, $\de \le O(k^2\log n+ k^3)$, so 
\begin{gather}  k^2\de \le O(k^4\logn+k^5). \label{eq:boundOPT}\end{gather}
By \Cref{treepacking}, the collection $\m T$ has size $k^{O(k)}n^{(1+o(1))k}$. For each $T\in\m T$, TreeCut$(G,T,k^2\de)$ is executed in line~\ref{line:treecut}, which runs in $k^{O(k)}(k^2\de)^k n^{o(k)}$ time by \Cref{thm:tree}.  In total, this is $k^{O(k)}n^{(1+o(1))k} \cd k^{O(k)}(k^2\de)^kn^{o(k)}$ time, which is at most $k^{O(k)}(\log n)^kn^{(1+o(1))k}$ by (\ref{eq:boundOPT}). The $(\logn)^k$ factor is negligible by \Cref{lognk}.

Otherwise,   line~\ref{line:m10} holds. By \Cref{thm:NI} with $s:=k^2\de$, the graph $H$ has at most $k^2\de n$ edges, so by \Cref{thm:KT}, the graph $G$ in line~\ref{line:m10} has $\ol n=\tO(k^2 (k^2\de)/\de^2)=\tO(k^{O(1)}n/\de)$ vertices. By the same arguments as in the previous paragraph, the calls to TreeCut$(G,T,k^2\de)$ take $k^{O(k)}(k^2\de)^k\ol n^{(1+o(1))k}$ time, which is bounded by $k^{O(k)}(k^2\de)^k (\tO(k^{O(1)}n/\de))^{(1+o(1))k} = k^{O(k)}n^{(1+o(1))k}$.

Finally, we handle the recursive component of the algorithm. Fix an arbitrarily small $\e>0$, and fix constants $c_1,c_2$ such that the nonrecursive part takes time $k^{c_1 k} n^{(1+\e)k + c_2}$ for all $k$. We prove by induction on $k$ that the total algorithm takes time $(2k)^{c_1k}n^{(1+\e)k+c_2}$, with the trivial base case $k=1$. For $k>1$, the $n$ recursive calls to $(k-1)$-cut take $n\cd(2 k)^{c_1(k-1)}n^{(1+\e)(k-1)+c_2} = (2k)^{-c_1}(2k)^{c_1k}n^{(1+\e)k+c_2}$ time total. The nonrecursive part takes time   $k^{c_1 k} n^{(1+\e)k + c_2} = 2^{-c_1k}(2k)^{c_1k}n^{(1+\e)k+c_2}$ by assumption. As long as $(2k)^{-c_1}+2^{-c_1k}\le1$, which holds for any constant $c_1\ge1$, the sum of the two running times is at most $(2k)^{c_1k}n^{(1+\e)k+c_2}$, preserving the induction. Hence, the running time is $k^{O(k)}n^{(1+o(1))k}$.
\EP

\section{Algorithm on Tight T-trees}\label{sec:tree}

In this section, we prove the running time guarantee of TreeCut$(G,T,s)$ in line~\ref{line:treecut}.

\tree*

Since the minimum $k$-cut can be obtained by deleting $k-1$ edges of $T$ and taking the connected components as the $k$-cut, our algorithm will pursue this route: it will look for the best $k-1$ edges of $T$ to delete. Let $E_T^*:=E_T[S^*_1,\lds,S^*_k]$ be the optimal set of $(k-1)$ edges to delete. 

First, observe that we can assume that for every edge $(u,v)$ in $T$, the minimum (2-)cut that separates $u$ and $v$ has size~$\le\la$. This is because if an edge $(u,v)$ in $T$ does not satisfy this property, then no minimum $k$-cut of size~$\le\la$ can separate $u$ and $v$, so we can contract $u$ and $v$ in $T$. Moreover, since $s$--$t$ minimum cut is polynomial time solvable, the algorithm can detect which edges to contract.
\BA\label{as:con}
For every edge $(u,v)$ in $T$, the minimum (2-)cut that separates $u$ and $v$ has size~$\le\la$.
\EA

\subsection{Restricted Case: Union of Branches}\label{sec:single}

We first begin with an algorithm when the tree $T$ is ``spider-like'', as discussed in \Cref{thm:tec}.

\BT
Let $G$ be an unweighted multigraph, let $T$ be a tight \Ttree of $G$, and let $\la$ be a parameter. Suppose in addition that:
\BE
\im We can root $T$ at a vertex $r\in V(T)$ so that $T$ is a disjoint union of maximal branches. 
\im There is an optimal minimum $k$-cut $S^*_1,\lds,S^*_k$ such that $E_T[S^*_1,\lds,S^*_k]$ contains at most one edge from each maximal branch.
\EE 
Then, there is an algorithm \textup{TreeCut}$(G,T,s)$ with the following guarantee: if the minimum $k$-cut in $G$ has size~$\le s$, then \textup{TreeCut} outputs a minimum $k$-cut of $G$. The running time of \textup{TreeCut} is $k^{O(k)}s^kn^{o(k)}$.
\ET

In this section, we develop an algorithm to solve this restricted case. Throughout, we assume that the minimum $k$-cut of $G$ is indeed at most $\la$, since otherwise, the algorithm can output anything.

Since the edges of $E^*_T$ lie in distinct branches rooted at the same vertex, the child vertices of the edges in $E^*_T$ are incomparable, and each subtree rooted at a child vertex is a component in $\m S$. Without loss of generality, let $S^*_k$ be the component containing the root of $T$.
 Let $v^*_1,\lds,v^*_{k-1}$ be the child vertices of $E^*_T$ such that component $S^*_i$ is exactly $T(v^*_i)$. For each $i\in[k-1]$, consider the maximal branch containing $v^*_i$, and let $u^*_i$ be the child of the root $r$ that lies on this branch. Lastly, define $E'\s E$ to be the edges whose endpoints are incomparable (i.e., they do not lie on a common branch).

\BD\label{def:precede}
An edge $(u,v)\in E(T)$ is \emph{partially preceded} by a vertex $x\in V(T)$ if either $x$ precedes $u$ or $x$ precedes $v$ (or both).
\ED

Define the multigraph $H$ as the graph obtained from starting with $G[\bigcup_{i\in[k-1]}V(T(v^*_i))]$ and contracting each vertex set $V(T(v^*_i))$ ($i\in[k-1]$) into a single vertex $v^*_i$, with self-loops removed. More precisely, $H$ has vertex set $\{v^*_1,\lds,v^*_{k-1}\}$, and its edge set is as follows: for each edge $e\in E'$ with endpoints in $T(v^*_i)$ and $T(v^*_j)$ ($i\ne j$), add that same edge $e$ between $v^*_i$ and $v^*_j$. Note that the two graphs share common vertices and edges; we make it this way to facilitate transitioning between the two graphs. Observe that $|E(H)|\le\la$, since every edge in $E(H)$ corresponds to an edge in $E_G(S^*_1,\lds,S^*_{k-1})$. 
 Moreover, for an edge $e\in E(H)$ with endpoints $v^*_i,v^*_j$, edge $e$ connects $T(v^*_i)$ and $T(v^*_j)$ in $G$; let $e|v^*_i$ and $e|v^*_j$ denote the endpoint of $e$ in $T(v^*_i)$ and $T(v^*_j)$, respectively.

\BL\label{lem:spanning-tree}
For each connected component $C$ in $H$, there exists a spanning tree $T_C$ of $C$ satisfying the following property: Let $U$ be the set of endpoints of edges in $T_C$ (more formally, $U:=\bigcup_{(u,v)\in E(T_C)}\{u,v\}$). Then, every edge $e$ in $C$ is partially preceded by some vertex in $U$ (in the tree $T$).
\EL
\BP

Fix a connected component $C$ of $H$, and construct a weighted digraph $H'$ as follows: for each edge $e\in E(C)$ with endpoints $v^*_i,v^*_j$, add an arc $(v^*_i,v^*_j)$ with weight equal to the depth of vertex  $e|v^*_j$ in $T(v^*_j)$, and an arc $(v^*_j,v^*_i)$ with weight equal to the depth of  vertex $e|v^*_i$ in $T(v^*_i)$. Since $C$ is connected, $H'$ is strongly connected. Let $r\in V(H')$ be arbitrary, and let $A\s E(H')$ be a minimum cost (out-)arborescence of $H'$ rooted at $r$ (so that $r$ is the only vertex with no in-arcs in $A$). We claim that the tree $\overline A$ formed by un-directing every arc in $A$ is our desired spanning tree $T_C$.

To prove this claim, let $e\in E(C)$ be arbitrary, with endpoints $v^*_i,v^*_j$, and assume without loss of generality that $v^*_i$ does not precede $v^*_j$ in the tree $\overline A$ rooted at $r$ (otherwise we can swap $v^*_i$ and $v^*_j$). Let $a\in E(H')$ be the arc originating from $e$ in the direction $(v^*_j,v^*_i)$, and let $a'\in H[C]$ be the in-arc of $v^*_i$ in $A$ (note that $v^*_i$ cannot be the root since it does not precede $v^*_j$, and that it is possible that $a=a'$). Observe that $A \setminus a'\cup a$ is also an arborescence. Since $A$ is the minimum cost arborescence, the weight of $a'$ is at most the weight of $a$. Let $e'$ be the edge originating from $a'$; we have $\depth( e'|v^*_i)\le\depth( e|v^*_i)$. It follows that the endpoint $ e|v^*_i$ is preceded by $ e'|v^*_i\in U$.
\EP

\subsubsection{Restricted Case: Algorithm}\label{sec:single-alg}

\begin{figure}
\begin{tikzpicture}

\node[left] at (-4,3) {$u^*_1$};
\node[left] at (-2,3) {$u^*_2$};
\node[] at (0,3) {$u^*_3 \ \ v^*_3$};
\node[right] at (2,3) {$u^*_4$};
\node[right] at (4,3) {$u^*_5$};
\node[right] at (6,3) {$u$};
\node[right] at (8,3) {$u'$};
\node[above] at (0,4) {$r$};

\node (v8) at (-6,-5) {};
\node (v9) at (-3,-5) {};
\node (v10) at (0,-5) {};
\node (v11) at (3,-5) {};
\node (v12) at (6,-5) {};
\node (v13) at (8,-5) {};

\node[cyan] at (-6.1618,-4.5602) {$S^*_1$};
\node[cyan] at (-3.3213,-4.8089) {$S^*_2$};
\node[cyan] at (-0.5463,-4.7042) {$S^*_3$};
\node [cyan]at (2.4251,-4.7827) {$S^*_4$};
\node[cyan] at (5.475,-4.6911) {$S^*_5$};

\node[left] (14) at (-5,-1) {$v^*_1$};
\node[right] (15) at (-2.6182,-2.0054) {$v^*_2$};
\node[right] at (2.5577,-1.4738) {$v^*_4$};
\node[left] at (4.6843,0.2315) {$v^*_5$};

\tikzstyle{every node}=[circle, fill, scale=.5];

\node (v2) at (0,4) {};
\node (v4) at (0,3) {};
\node (v3) at (-2,3) {};
\node (v1) at (-4,3) {};
\node (v5) at (2,3) {};
\node (v6) at (4,3) {};
\node (v7) at (6,3) {};

\node (v14) at (-5,-1) {};
\node (v15) at (-2.6182,-2.0054) {};
\node at (2.5577,-1.4738) {};
\node at (4.6843,0.2315) {};
\node (v16) at (4.6851,0.2296) {};
\node (v17) at (8,3) {};

\tikzstyle{every node}=[];

\draw [cyan , line width=1] plot[smooth, tension=.7] coordinates {(-6.8608,-4.7527) (-5.5964,-1.0129) (-4.4803,-1.4917) (-5.0307,-5.0004)};
\draw [cyan , line width=1] plot[smooth, tension=.7] coordinates {(-3.7924,-4.9729) (-2.8999,-2.0077) (-1.9967,-2.1909) (-2.1274,-5.0004)};
\draw  [cyan , line width=1]plot[smooth, tension=.7] coordinates {(-1.5124,-5.0314) (-0.8212,2.3512) (0.8831,2.3512) (1.4066,-4.9922)};
\draw [cyan , line width=1] plot[smooth, tension=.7] coordinates {(1.8629,-5.0417) (2.1397,-1.7197) (3.2654,-1.5757) (3.9406,-4.8766)};
\draw  [cyan , line width=1]plot[smooth, tension=.7] coordinates {(5.0277,-5.0417) (4.077,-0.0966) (5.3854,-0.2808) (6.9815,-4.9729)};
\draw[red,dashed , line width=1] (v14) -- (-2.4597,-0.7847);

\draw[red , line width=1] (-5.3956,-2.6565) -- (-2.8266,-3.6938);
\draw[green , line width=1] (-5.6908,-3.8449) -- (v15);
\draw[green , line width=1] (-2.949,-4.6682) -- (0.0024,0.9008);
\draw[red,dashed , line width=1] (-2.2112,1.2581) -- (0.0102,0.2018);
\draw[red ,dashed, line width=1] (-2.351,0.1707) -- (0.0102,-4.008);
\draw[red , line width=1] (-0.0053,-3.4487) -- (2.8995,-4.3031);
\draw[red , line width=1] (2.7753,-3.1303) -- (4.8413,-0.4118);
\draw[red,dashed , line width=1] (2.4102,-0.3031) node (v18) {} -- (4.8413,-0.4118);

\draw[green , line width=1] (-2.8708,-4.2253) -- (5.1305,-1.6566);
\draw[green , line width=1] (5.6131,-3.5091) --  (2.7753,-3.1303);

\draw[red,dashed , line width=1] (v16) -- (6.4557,1.046);
\draw[red,dashed , line width=1] (5.353,-2.4953) -- (7.1873,-1.8909);
\draw[red,dashed , line width=1] (0.0199,1.5762) -- (2.1828,1.502);

\draw[red,dashed , line width=1] (-4.2525,1.912) -- (-2.1194,2.1059);
\draw [red,dashed , line width=1](2.2924,0.5868) -- (v6);

\draw[line width=1.5] (v8) -- (v1) -- (v2) -- (v3) -- (v9) -- (v3) -- (v2) -- (v4) -- (v10) -- (v4) -- (v2) -- (v5) -- (v11) -- (v5) -- (v2) -- (v6) -- (v12) -- (v6) -- (v2) -- (v7) -- (v13);

\draw [gray , line width=1]  plot[smooth, tension=.7] coordinates {(-4.4314,1.0946) (-4.3921,-0.3846) (-5.1644,-1.8113)};
\draw  [gray , line width=1] plot[smooth, tension=.7] coordinates {(6.2106,2.1287) (6.0666,0.5841) (6.8782,-0.6594)};
\draw  [gray , line width=1] plot[smooth, tension=.7] coordinates {(0.0322,-0.5024) (0.4118,-1.6412) (0.006,-2.8062)};

\draw[line width=1.5] (v2) -- (v17) -- (10,-5);
\draw [gray , line width=1] (7.4672,-2.924) -- (8.6191,0.4925);
\draw[gray , line width=1]  (6.6557,0.3616) -- (9.0904,-1.4448);
\draw  [gray , line width=1] plot[smooth, tension=.7] coordinates {(2.0661,2.3806) (1.7591,1.007) (v18)};
\end{tikzpicture}
\caption{In this example, the graph $H$ is connected, so there is only one component $C^*_1$. The green edges form $T^*_1$ (see \Cref{cond:color-i}); the (solid and dashed) red edges form $\bigcup_i\pt_G(T(u^*_i))\setminus\bigcup_iT^*_i$ (see \Cref{cond:color-0}); the solid red and green edges form $E[S^*_1,\lds,S^*_{k-1}]$; the gray edges can be either red or green without affecting Conditions~\ref{cond:color-i}~or~\ref{cond:color-0}. (Note that $E[S^*_1,\lds,S^*_5]$ is not \emph{actually} the minimum $6$-cut in the graph, but that is not the focus of this example.)}
\label{f1}
\end{figure}
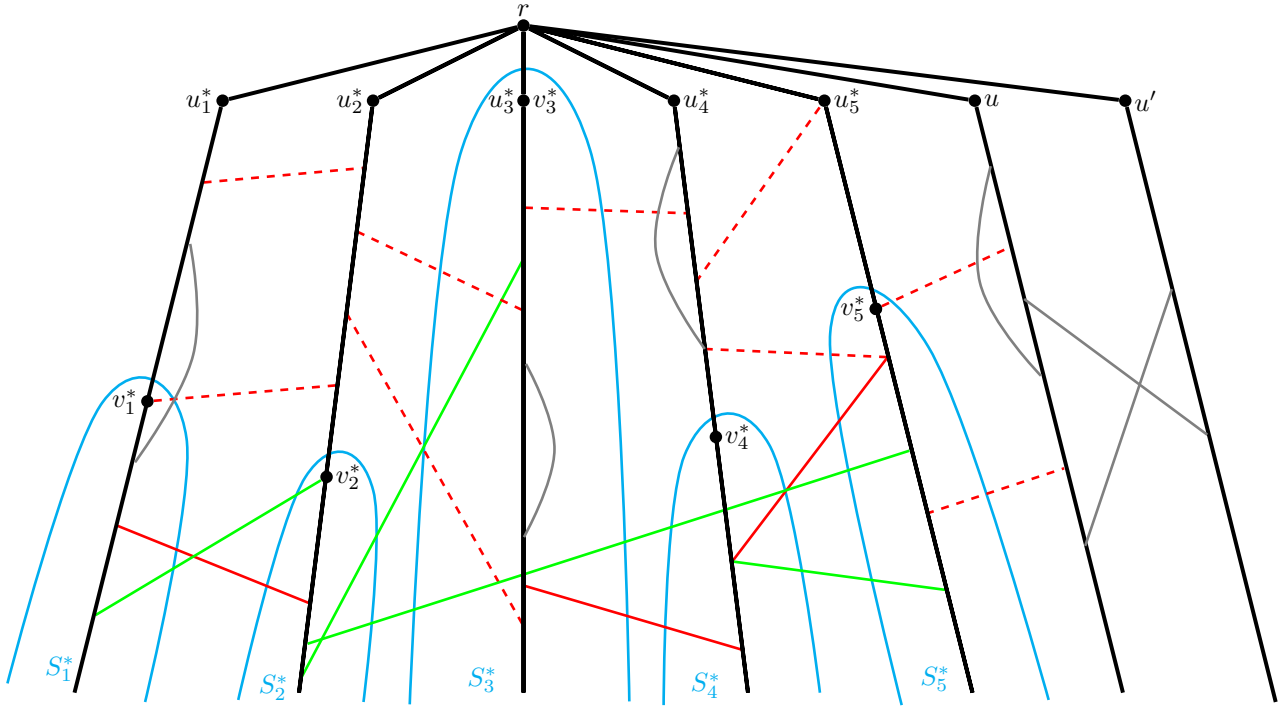

We first present the main steps our algorithm. Suppose for simplicity that $H$ is connected; that is, there is only one connected component $H$. In fact, \textbf{we encourage the reader to assume that $H$ is connected on their first reading}, since it simplifies the presentation while still preserving all the key insights. Our first insight is \emph{color-coding} to mark out the spanning tree $T_H$ guaranteed by \Cref{lem:spanning-tree}. In particular, all the edges in the spanning tree should be colored one color (say, green), while all other edges in $\pt_G(T(u^*_i))$ should be colored a different color (say, red); see Figure~\ref{f1}. This color-coding process will succeed with probability roughly $\la^{-k}$, so we need to repeat it roughly $\la^k$ many times. This is where we pay the $\la^k$ multiplicative factor in the running time. Then, construct a graph with the maximal branches as vertices, where two vertices are connected by an edge if their corresponding branches have a green edge between them. Assuming that $H$ is connected, one of these connected components corresponds exactly to the branches $T(u^*_1),\lds,T(u^*_{k-1})$. Finally, we iterate over the connected components $C$ of size $k-1$ (one of which captures $T(u^*_1),\lds,T(u^*_{k-1})$) and, with the information of the green edges, compute an overestimate of the minimum possible $k$-cut formed by cutting one edge from each of the corresponding $k-1$ branches of $C$. The catch is that for the component containing $T(u^*_1),\lds,T(u^*_{k-1})$, this estimate will actually be \emph{exact}. This ensures that the minimum $k$-cut is indeed returned.

The key insight in the algorithm is coloring the edges of this tree $T_H$, which serves two purposes. First, it allows the algorithm to figure out which $k-1$ branches contain $T_H$ by computing the connected components as described above. Second, the edges of $T_H$ partially precede all edges in $H$, including all edges that appear \emph{twice} in $\pt_G T(v^*_1),\lds,\pt_G T(v^*_{k-1})$. It turns out that these edges are the hardest to deal with, since they are the ones double-counted when merely summing up the boundaries $\pt_G T(v^*_1),\lds,\pt_G T(v^*_{k-1})$.\footnote{Indeed, a reader familiar with the $k$-Partial~Vertex~Cover problem may be familiar with the difficulty of double-counting, an complication that alone justifies why the problem is $W[1]$-hard.} However, with the knowledge of $T_H$, any edge partially preceded by an endpoint in $T_H$ is cut for sure, and this includes all double-counted edges! And once the double-counted edges are dealt with, we can simply treat the $k-1$ branches independently.\footnote{An illustrative analogy for the  minimum/maximum $k$-Partial Vertex Cover problem is that if we somehow knew that there were \emph{no} edges between the $k$ optimal vertices to select, then the problem becomes easy: simply output the $k$ vertices of minimum/maximum degree.}

It turns out that  both of these properties---finding the $k-1$ branches and dealing with the double-counted edges---can each be done separately with a $\la^k$ multiplicative factor, each with standard color-coding techniques.\footnote{Indeed, the $(1+\e)$-approximate $k$-cut algorithm time of \cite{GLL18b} can be adapted this way to solve exact minimum $k$-cut.} Focusing on this special tree $T_H$ is what enables us to achieve both with just one $\la^k$ factor.

We now proceed to the algorithm. Let the connected components of $H$ be $C^*_1,\lds,C^*_z$ for $z\le V(H)=r$, ordered in an arbitrary order, and let $T^*_i$ be the spanning tree for $C^*_i$ promised by \Cref{lem:spanning-tree}. The algorithm now colors the edges of $E'$ \red and \green such that the following two conditions hold:
\BCO\label{cond:color-i}
For each $C^*_i$, all edges in $T^*_i$ are colored \green.
\ECO
\BCO\label{cond:color-0}
All edges in $\bigcup_i\pt_G(T(u^*_i))\setminus \bigcup_{i}T^*_i$ are colored \red.
\ECO
We do this with the following color-coding procedure: color each edge \red with probability $1-1/\la$ and \green with probability $1/\la$. For each $i$, $\pt_G(T(u^*_i))$ is the ($2$-)cut in $G$ formed by removing the parent edge of $u^*_i$, so by \Cref{as:con}, we have $|\pt_G(T(u^*_i))|\le\la$. Therefore, the success probability is at least
\begin{gather} \lp1-\f1\la\rp^{\left|\bigcup_i\pt_G(T(u^*_i))\right|} \cd \Prod_{i\in[z]}\lp\f1{\la}\rp^{|E(T^*_i)|} \ge \lp1-\f1\la\rp^{k\la} \cd \lp\f1{\la}\rp^k = 2^{-O(k)}\la^{-k} . \label{eq:color}  \end{gather}
We describe this algorithm and its guarantees succinctly as follows:
\BAL\label{alg:color}
Color each edge in $E'$ \red with probability $1-1/\la$ and \green with probability $1/\la$. With probability $2^{-O(k)}\la^{-k}$, \Cref{cond:color-i,cond:color-0} hold.
\EAL

Next, build a graph whose vertices are the children of $r$, and for every two children $u,u'$ of $r$, connect them by an edge if there is a \green edge between $T(u)$ and $T(u')$. Consider all (maximal) connected components in this graph; for each one, add its set $U$ of vertices into a collection $\m U$.
 Observe that if \Cref{cond:color-i,cond:color-0} hold, then for each connected component $C^*_i$, there exists a set $U^*_i\in\m U$ of  size $|V(C^*_i)|$  such that each vertex $v^*_j$ in $C^*_i$ belongs on a (different, unique) branch $T(u)$ ($u\in U^*_i$). 
Therefore, with a success probability of $2^{-O(k)}\la^{-k}$, we can assume the following:
\BA\label{as:good}
After running \Cref{alg:color}, \Cref{cond:color-i,cond:color-0} hold, and for each connected component $C^*_i$, there exists a set $U^*_i\in\m U$ of  size $|V(C^*_i)|$  such that each vertex $v^*_j$ in $C^*_i$ belongs on a (different, unique) branch $T(u)$ ($u\in U$).  (This assumption holds with probability at least $2^{-O(k)}\la^{-k}$.)
\EA
We will not assume \Cref{as:good} unconditionally throughout the remainder of this section; rather, we will explicitly state where \Cref{as:good} is assumed. Hopefully, this provides more intuition as to how \Cref{as:good} is used.

\subsubsection{Restricted case: Processing the sets $U\in\m U$}\label{sec:mincut-setup}

In this section, we process the sets $U\in\m U$, solving a certain cut problem on a specific graph for each $U$ (in polynomial time). Recall that if $H$ is connected, then this is the step where we should compute the exact value $|OPT|$ for the corresponding vertex set $U^*_1=V(H)$. In general, for each set $U^*_i$ corresponding to a component $C^*_i$ of $H$ with vertices $v^*_{i_1},\lds,v^*_{i_\el}\in V^*_T$, we want to compute the exact value $|E[S^*_{i_1},\lds,S^*_{i_\el}]|$, whose sum over all $C^*_i$ will turn out to equal $|OPT|$. Moreover, for every other set $A\in\m A$, we want to make sure we compute some sort of \emph{overestimate}, so that these extraneous sets do not mislead the algorithm.

For each $U\in\m U$, define $\minelts(U) :=  \big(\bigcup_{u\in U}\bigcup_{(v,v')\in \pt T(u_i)\text{ \green}}(\{v,v'\}\cap T(u))\big) {\downarrow}$ as the minimal elements on $\bigcup_uT(u)$ of the set of endpoints of \green edges in $\bigcup_i\pt T(u_i)$, which is clearly pairwise incomparable. We now define the \emph{minimum ancestor cut} problem:
\BD[Minimum ancestor cut]
Fix some set $U=\{u_1,\lds,u_\el\}\in\m U$, and let $\minelts(U)=\{s_1,\lds,s_\el\}$ where $s_i\in V(T(u_i))$. The \emph{minimum ancestor cut} is the following problem: For each $i\in[\el]$, select one edge in the branch from $s_i$ to $r$, and consider the $(\el+1)$-cut in $G$ formed by removing these selected edges in $T$. We want to compute the $(\el+1)$-cut of minimum size, denoted $\minanccut(U)$.
\ED

Why are minimum ancestor cuts relevant? We first show that for components $C^*_i$ of $H$, $\minanccut(U^*_i)$ has a close connection with $S^*_1,\lds,S^*_k$.
\BCL\label{clm:anc}
Assuming \Cref{as:good}, for each  $ U^*_i=\{u^*_{i_1},\lds,u^*_{i_\el}\}$, we have $|\pt_G[S^*_{i_1},\lds,S^*_{i_\el}]|=|\minanccut(U^*_i)|$. Moreover, $\sum_{U^*_i}|\minanccut(U^*_i)|=OPT$.
\ECL
\BP
Recall that the corresponding component $C^*_i$ has vertices $v^*_{i_1},\lds,v^*_{i_\el}$. First, we claim that an ancestor cut of size $|\pt_G[S^*_{i_1},\lds,S^*_{i_\el}]|$ is achievable: simply cut the parent edges of $v^*_{i_1},\lds,v^*_{i_\el}$. We now show that this cut is indeed an ancestor cut. By \Cref{cond:color-i,cond:color-0}, the only \green edges in any $\pt T(u^*_{i_j})$ lie in $T^*_i$. Since $T^*_i\s H$, and since any edge in $H$ between $T(u^*_{i_j})$ and $T(u^*_{i_{j'}})$ must lie between $T(v^*_{i_j})$ and $T(v^*_{i_{j'}})$, we have that $v^*_{i_j}$ precedes any endpoint of a \green edge on $T(u^*_{i_j})$. Therefore, the parent edges of $v^*_{i_j}$ lie on the branches from endpoints in $\minelts(U^*_i)$ to $r$, so our cut is a valid ancestor cut.

We now show that no better ancestor cut is possible; suppose otherwise. Then, let $S_{i_1},\lds,S_{i_\el}$ be the components not containing $r$ in $\minanccut(U^*_i)$. First, merge the components $S^*_{i_1},\lds,S^*_{i_\el}$ together with $S^*_k$. We claim that the number of cut edges drops by exactly $|\pt_G[S^*_{i_1},\lds,S^*_{i_\el}]|$: since $C^*_i$ is a component of $H$, all edges in $\pt_G[S^*_{i_1},\lds,S^*_{i_\el}]$ have endpoints in $S^*_{i_1},\lds,S^*_{i_\el}$ or $S^*_k$ (instead of outside these components). Next, split $S^*_{i_1}\cup\lds\cup S^*_{i_\el}\cup S^*_k$ into $S_{i_1},\lds,S_{i_\el}$ and the remaining component; this increases the number of cut edges by at most $|\minanccut(U^*_i)|$. We arrive at a $k$-cut of smaller size, contradicting the choice of $S^*_1,\lds,S^*_k$.

Finally, to prove that $\sum_{U^*_i}|\minanccut(U^*_i)|=|\pt_G[S^*_1,\lds,S^*_k]|$,
observe that over different components $C^*_j$ of $H$ with vertices $v^*_{j_1},\lds,v^*_{j_\el}\in V^*_T$, the edges $\pt_G[S^*_{j_1},\lds,S^*_{j_\el}]$ are disjoint over distinct $C^*_j$. (This is because if two distinct $C^*_j,C^*_{j'}$ shared an edge in their respective edge sets $\pt_G[\cd,\lds,\cd]$, then the components $C^*_j,C^*_{j'}$ should have become a single connected component.) Therefore, $|\pt_G[S^*_1,\lds,S^*_k]|$ equals the sum of the $|\pt_G[S^*_{j_1},\lds,S^*_{j_\el}]|$ values, which equals the sum of the $|\minanccut(U^*_i)|$ values.
\EP

Therefore, for each $U^*_i$, we may define $\minanccut(U^*_i)$ as not just any arbitrary minimum ancestor cut, but the specific one formed by cutting the edges $E_T[S^*_{i_1},\lds,S^*_{i_\el}]$. (By \Cref{clm:anc}, this is a minimum ancestor cut.)


We now compute a function $f(U)$ for each set $U\in\m U$. For each $U^*_i$, we want $f(U^*_i)=|\minanccut(U^*_i)|$ so that it exactly captures the contribution of $U^*_i$ to the minimum $k$-cut. (We will need \Cref{as:good} to achieve this.) For any other $U$, we want $f(U)\ge|\minanccut(U)|$. Lastly, we want $f(U)$ to be computable in polynomial time.

\begin{figure}
\begin{tikzpicture}

\node[left] at (-4,3) {$u_1$};
\node[left] at (-2,3) {$u_2$};
\node[] at (0,3) {$u_3 \ \ v^*_3$};
\node[right] at (2,3) {$u_4$};
\node[right] at (4,3) {$u_5$};
\node[above] at (0,4) {$r$};
\node[left] at (-5.7109,-3.8487) {$s_1$};
\node[left] at  (-2.6282,-2.0054) {$s_2$};
\node[right] at (0.0024,0.9008) {$s_3$};
\node[left] at (2.7666,-3.1293) {$s_4$};
\node[right] at (5.1573,-1.6316) {$s_5$};

\node (v8) at (-6,-5) {};
\node (v9) at (-3,-5) {};
\node (v10) at (0,-5) {};
\node (v11) at (3,-5) {};
\node (v12) at (6,-5) {};

\node[cyan] at (-6.1618,-4.5602) {$S^*_1$};
\node[cyan] at (-3.3213,-4.8089) {$S^*_2$};
\node[cyan] at (-0.5463,-4.7042) {$S^*_3$};
\node [cyan]at (2.4251,-4.7827) {$S^*_4$};
\node[cyan] at (5.475,-4.6911) {$S^*_5$};

\node[left] (14) at (-5,-1) {$v^*_1$};
\node[right] (15) at (-2.6182,-2.0054) {$v^*_2$};
\node[right] at (2.5577,-1.4738) {$v^*_4$};
\node[left] at (4.6843,0.2315) {$v^*_5$};

\tikzstyle{every node}=[circle, fill, scale=.5];

\node (v2) at (0,4) {};
\node (v4) at (0,3) {};
\node (v3) at (-2,3) {};
\node (v1) at (-4,3) {};
\node (v5) at (2,3) {};
\node (v6) at (4,3) {};
\node at (-5.7109,-3.8487) {};
\node at  (-2.6282,-2.0054) {};
\node at (0.0024,0.9008) {};;
\node at (2.7666,-3.1293) {};
\node at (5.1573,-1.6316) {};

\node (v14) at (-5,-1) {};
\node (v15) at (-2.6282,-2.0054) {};
\node at (2.5577,-1.4738) {};
\node at (4.6843,0.2315) {};
\node (v16) at (4.6851,0.2296) {};

\tikzstyle{every node}=[];

\draw [cyan , line width=1] plot[smooth, tension=.7] coordinates {(-6.8608,-4.7527) (-5.5964,-1.0129) (-4.4803,-1.4917) (-5.0307,-5.0004)};
\draw [cyan , line width=1] plot[smooth, tension=.7] coordinates {(-3.7924,-4.9729) (-3.0819,-1.8795) (-1.9967,-2.1909) (-2.1274,-5.0004)};
\draw  [cyan , line width=1]plot[smooth, tension=.7] coordinates {(-1.5124,-5.0314) (-0.8212,2.3512) (0.8831,2.3512) (1.4066,-4.9922)};
\draw [cyan , line width=1] plot[smooth, tension=.7] coordinates {(1.8629,-5.0417) (2.1397,-1.7197) (3.2654,-1.5757) (3.9406,-4.8766)};
\draw  [cyan , line width=1]plot[smooth, tension=.7] coordinates {(5.0277,-5.0417) (4.077,-0.0966) (5.3854,-0.2808) (6.9815,-4.9729)};
\draw[purple,dashed , line width=1] (v14) -- (-2.4597,-0.7847) node (v25) {};

\draw[brown ,dashed  , line width=1] (-5.3956,-2.6565) node (v24) {} -- (-2.8266,-3.6938) node (v26) {};
\draw[brown ,dashed  , line width=1] (-5.7109,-3.8487) node (v7) {} -- (v15);
\draw[brown ,dashed  , line width=1] (-2.949,-4.6682) -- (0.0024,0.9008) node (v17) {};
\draw[purple,dashed , line width=1] (-2.2112,1.2581) node (v23) {} -- (0.0102,0.2018) node (v27) {};
\draw[brown ,dashed, line width=1] (-2.351,0.1707) -- (0.0102,-4.008);
\draw[brown ,dashed  , line width=1] (-0.0053,-3.4487) -- (2.8995,-4.3031);
\draw[brown ,dashed  , line width=1] (2.7666,-3.1293) -- (4.8413,-0.4118);
\draw[purple,dashed , line width=1] (2.4102,-0.3031) node (v18) {} -- (4.8413,-0.4118) node (v31) {};

\draw[brown ,dashed  , line width=1] (-2.8993,-4.2376) node (v13) {} -- (5.1573,-1.6316) node (v20) {};
\draw[brown ,dashed  , line width=1] (5.6131,-3.5091) --  (2.7666,-3.1293) node (v19) {};

\draw[purple ,dashed , line width=1] (0.0199,1.5762) node (v28) {} -- (2.1828,1.502) node (v29) {};

\draw[purple ,dashed , line width=1] (-4.2525,1.912) node (v21) {} -- (-2.1194,2.1059) node (v22) {};
\draw [purple,dashed , line width=1](2.2924,0.5868) node (v30) {} -- (v6);

\draw[line width=1.5] (v8) -- (v1) -- (v2) -- (v3) -- (v9) -- (v3) -- (v2) -- (v4) -- (v10) -- (v4) -- (v2) -- (v5) -- (v11) -- (v5) -- (v2) -- (v6) -- (v12) -- (v6) -- (v2);

\draw [blue ,  line width=1.5]  plot[smooth, tension=.7] coordinates {(-4.4314,1.0946) (-4.3921,-0.3846) (-5.1644,-1.8113)};

\draw  [orange ,dashed  , line width=1] plot[smooth, tension=.7] coordinates {(0.0322,-0.5024) (0.4118,-1.6412) (0.006,-2.8062)};

\draw  [blue,  line width=.7] plot[smooth, tension=.7] coordinates {(2.0661,2.3806) (1.7591,1.007) (v18)};

\draw[green,line width=2]  (v7.center) edge (v8);
\draw [green,line width=2]   (v15.center) edge (v9);
\draw [green,line width=2]   (v17.center) edge (v10);
\draw [green,line width=2]   (v19.center) edge (v11);
\draw [green,line width=2]   (v20.center) edge (v12);

\draw[purple,line width=.7]  plot[smooth, tension=.7] coordinates {(v21) (-2.95,3.0407) (v2)};
\draw[purple,line width=.7]  plot[smooth, tension=.7] coordinates {(v22) (-1.136,3.1909) (v2)};
\draw[purple,line width=.7]  plot[smooth, tension=.7] coordinates {(v23) (-1.2054,2.798) (v2)};
\draw [purple,line width=1.5] plot[smooth, tension=.7] coordinates {(v14) (-3.2042,2.5323) (v2)};
\draw [purple,line width=.7] plot[smooth, tension=.7] coordinates {(v25) (-1.448,1.9546) (v2)};
\draw [purple,line width=.7] plot[smooth, tension=.7] coordinates {(v13)};
\draw [purple,line width=1.5] plot[smooth, tension=.7] coordinates {(v27) (-0.281,2.2203) (v2)};
\draw [purple,line width=1.5] plot[smooth, tension=.7] coordinates {(v28) (0.2389,2.8096) (v2)};
\draw [purple,line width=.7] plot[smooth, tension=.7] coordinates {(v29) (1.5445,2.902) (v2)};
\draw [purple,line width=.7] plot[smooth, tension=.7] coordinates {(v30) (1.3943,2.7056) (v2)};
\draw [purple,line width=.7] plot[smooth, tension=.7] coordinates {(v18) (1.2441,2.4629) (v2)};
\draw [purple,line width=.7] plot[smooth, tension=.7] coordinates {(v6) (2.5959,3.214) (v2)};

\draw [purple,line width=1.5] plot[smooth, tension=.7] coordinates {(v31) (3.3123,2.3012) (v2)};
\end{tikzpicture}
\caption{The graph $G'$ for $U^*_1$ from Figure~\ref{f1}. The green branches are the vertices in $T(\minelts(U^*_1))$. The dotted orange, brown, blue, and purple edges are the edges in $E$ considered in step (a), (b), (c), and (d), respectively. The solid blue and purple edges are the edges in $G'$. The bold blue and purple edges are the ones cut in $\minanccut(U^*_1)$, which is also the cut which produces $S^*_1,\lds,S^*_5$.}
\label{f2}
\end{figure}
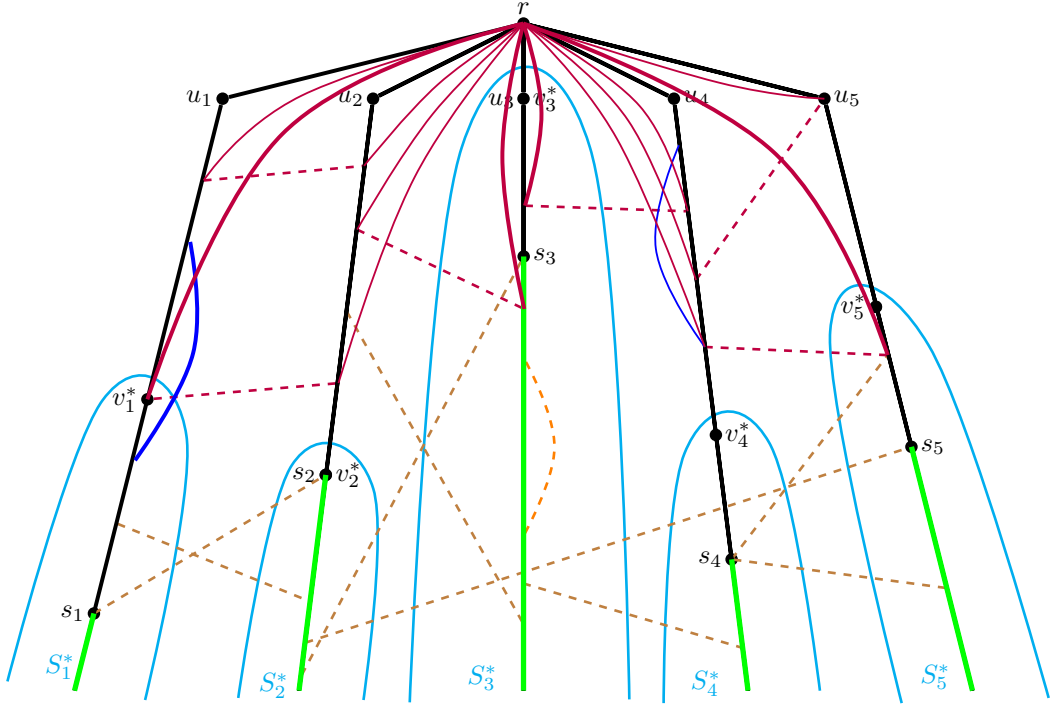

Fix a set $U=\{u_1,\lds ,u_\el\}$ and let $\minelts(U)=\{s_1,\lds,s_\el\}$ where $s_i\in V(T(u_i))$. To compute $f(U)$, we construct the following multigraph $G'$ on the vertices $\lp\bigcup V(T(u_i))\rp\cup r$ (see Figure~\ref{f2}):

\BE 
\im[(a)] For each edge in $E\setminus E'$ \footnote{ Recall that $E'\s E$ is the edges with incomparable endpoints in $T$.} with both endpoints in $T(\minelts(U))$, do nothing: these edges are not cut in any ancestor cut.
\im[(b)] For each edge in $E'$ with at least one endpoint in $T(\minelts(U))$, do nothing: these edges are \emph{always} cut in an ancestor cut, and we will account for these edges separately. 
\im[(c)] For each edge in $E\setminus E'$ with at most one endpoint in $T(\minelts(U))$, add it to $G'$.
\im[(d)] For each edge $(u,v)\in E'$ with both endpoints not in $T(\minelts(U))$, add edges $(u,r)$ and $(v,r)$ to $G'$.
\EE
Note that all edges in $G'$ have both their endpoints in the same branch. Now, for each $i\in[\el]$, compute the vertex $t_i\in T(u_i)$ that minimizes $|\pt_{G'}T(t_i)|$. Take the sum of the costs of these $\el$ cuts, and finally, add the number of edges considered in step (b) to this sum. The final value is $f(U)$.


\BAL\label{alg:mincut}
For each set $U\in\m U$, construct the graph $G'$ as above, and compute the vertex $t_i\in T(u_i)$ that minimizes $|\pt_{G'}T(t_i)|$. Take the sum of the costs of these cuts, and add the number of edges considered in step (b) to this sum. Let $f(U)$ be the final value.
\EAL

Let us now explain the intuition of the construction of $G'$, relating it to $\minanccut(U)$. First, every edge considered in (a) has both endpoints on the same maximal branch, both of which are below $s_i$ on the appropriate branch, so it is never cut in an ancestor cut and can therefore be ignored. Every edge considered in (b) is always cut in an ancestor cut: if the edge is $(u,v)$ with $u\in T(s_i)$, then $u$ and $s_i$ will always belong in the same component in the ancestor cut, but never $v$ and $s_i$ because $v$ is on a different branch. Every edge in (c) can either be cut or not cut depending on the specific ancestor cut, and it is easy to see that it is included in $\bigcup\pt_{G'}T(t_i)$ iff it is cut in the ancestor cut that cuts the parent edges of each $t_i$. Finally, every edge in (d) splits into two edges, possibly adding two edges to a cut in $G'$. Indeed, if neither endpoint of edge $(u,v)$ in (d) is in $r$'s side of the cut, then both corresponding edges in $G'$ are cut. This is where the overestimate $f(U)\ge|\minanccut(U)| $ will come from. However, assuming \Cref{as:good},  there cannot be any overestimate for each $U^*_i$: by definition of $T^*_i$, every edge in $E(S^*_1,\lds,S^*_{k-1})$ with an endpoint (equivalently, both endpoints) in the maximal branches containing vertices in $\minelts(U^*_i)$ is partially preceded by $\minelts(U^*_i)$. Therefore, we have equality for each $U^*_i$: $f(U^*_i)=|\minanccut(U^*_i)| $. 

We now formalize our intuition. Define the following natural correspondence between ancestor cuts and the ``$G'$-cuts'' $\bigcup_i\pt_{G'}T(t_i)$ over the choices of $\{t_i:i\in[\el]\}$: two correspond to each other if the edges in the ancestor cut are the same as the parent edges of $t_i$.
\BL\label{lem:upper-bound}
For two corresponding cuts, the size of the ancestor cut is at most the size of the $G'$-cut plus the number of edges in step~(b).
\EL
\BP
The proof essentially following the intuition paragraph above. An edge in (a) contributes $0$ to the sizes of both cuts, and an edge in (b) contributes $1$ to the ancestor cut and 1 to the number of edges in step~(b). An edge in (c) contributes an equal amount to both cuts. Finally, if an edge in (d) contributes $0$ to the ancestor cut, then both of its endpoints belong to the component containing $r$, so in the $G'$-cut, neither of its endpoints is in their respective $T(t_i)$; hence, it also contributes $0$ to the $G'$-cut. If the edge contributes $1$ to the ancestor cut, then it contributes either $1$ or $2$ to the $G'$-cut depending on whether exactly one endpoint belongs to $r$'s component in the ancestor cut ($1$ to $G'$-cut), or no endpoints belong to it ($2$ to $G'$-cut).
\EP
\BL\label{lem:equality}
Assuming \Cref{as:good}, for each $U^*_i$,  $\minanccut(U^*_i)$ has size exactly $f(U^*_i)$.
\EL
\BP
In the proof of \Cref{lem:upper-bound}, the only potential source of inequality is in (d): an edge with no endpoints in $r$'s component of $\minanccut(U^*_i)$ contributes $1$ to $|\minanccut(U^*_i)|$ and $2$ to the corresponding $G'$-cut. If such an edge $e$ existed, then it must be in $\pt_G[S^*_{i_1},\lds,S^*_{i_\el}]$ where $V(C^*_i):=\{v^*_{i_1},\lds,v^*_{i_\el}\}$, which means $e$ is in component $C^*_i$. Also, neither of its endpoints is preceded by a vertex in $\minelts(U^*_i)$, which means neither of its endpoints is preceded by any endpoint in $G$ of any edge in $C^*_i$, and therefore any endpoint in $G$ of any edge in $T^*_i$ as  well. We thus have an edge $e$ in $C^*_i$ not partially preceded by any endpoint in $T^*_i$, contradicting the definition of $T^*_i$ (see beginning of \Cref{sec:single-alg}). Therefore, no such edge exists, and we have equality.
\EP

We run \Cref{alg:mincut}, computing the value $f(U)$ for each set $U\in\m U$ in polynomial time. Finally, the algorithm seeks to minimize
\begin{gather}
\min_{\substack{U_1,\lds,U_\el \\ \sum_i|U_i|=k-1}} \sum_{i=1}^\el f(U_i) . \label{eq:sum-f}
\end{gather}
The expression (\ref{eq:sum-f}) can be formulated as a knapsack problem with small, integral costs, which can easily be solved in polynomial time.

Since the branches in any two distinct $U,U'$ are disjoint, the sum $\sum_if(U_i)$ for any $U_1,\lds,U_\el$ is a $\lp\sum_i|U_i|\rp$-cut. This fact, along with \Cref{lem:upper-bound}, proves that (\ref{eq:sum-f}) is at least $OPT$. Furthermore, assuming \Cref{as:good}, $OPT$ can be achieved by \Cref{lem:equality}. Thus, as long as \Cref{as:good} is true, (\ref{eq:sum-f}) is exactly $OPT$.
\BAL\label{alg:knapsack}
Compute (\ref{eq:sum-f}) in polynomial time by formulating it as a knapsack problem. Assuming \Cref{as:good}, the result is exactly $OPT$.
\EAL

\BL\label{lem:total-at-least}
(\ref{eq:sum-f}) is always at least $OPT$.
\EL
\BP
Let $U_1,\lds,U_{\el}$ be the sets achieving the minimum in (\ref{eq:sum-f}). For each $U_i$, consider the best $G'$-cut in the graph $G'$ constructed for $U_i$. By \Cref{lem:upper-bound}, the corresponding ancestor cut has size at most $f(U_i)$. Also, the ancestor cut has one edge sharing a maximal branch with each vertex in $\minelts(U_i)$. Since $\minelts(U_i)$ and $\minelts(U_j)$ lie on different branches for $i\ne j$, if we take the union of the ancestor cuts over all $U_i$, then every maximal branch is cut at most once, so we cut one edge from each of exactly $k-1$ maximal branches. This union is a $k$-cut of cost at most $\sum_if(U_i)$ (it could be smaller if an edge appears twice in the union, once from each side), which means that $\sum_if(U_i)\ge OPT$.
\EP

\BL\label{lem:final-single}
Assuming \Cref{as:good}, (\ref{eq:sum-f}) equals $OPT$.
\EL
\BP
By \Cref{lem:total-at-least}, it suffices to show that (\ref{eq:sum-f}) is at most $OPT$ (assuming \Cref{as:good}). By \Cref{lem:equality}, $|\minanccut(U^*_i)|=f(U^*_i)$ for each $i$. Finally, by \Cref{clm:anc},  $\sum_{U^*_i}|\minanccut(U^*_i)|=OPT$.
 Thus, $\sum_if(U^*_i)=OPT$, and the minimum in (\ref{eq:sum-f}) can only be smaller.
\EP

Thus, by \Cref{lem:final-single}, the algorithm below outputs an optimum $k$-cut w.h.p., proving THM-MAIN.
\BAL\label{alg:final-single}
For $O(2^{O(k)}\la^k\logn)$ repetitions, run \Cref{alg:color,alg:mincut,alg:knapsack}, and output the minimum value of (\ref{eq:sum-f}) ever computed.
\EAL

\subsection{General Case}

In this section, we present our general algorithm, proving \Cref{thm:tree}, restated below.
\tree*

Before we begin, let us briefly describe the differences of the general setting and state the techniques we will use to overcome the new difficulties.

\BE
\im The first difference is that in general, the $k-1$ edges in $E_T[S^*_1,\lds,S^*_k]$ may not be incomparable. For example, in the extreme case, they can all lie on a single maximal branch. We resolve this issue with \emph{dynamic programming} on the tree, in a similar fashion to Section~2.3.2 of \cite{GLL18b}. At a high level, we only focus on the \emph{minimal} vertices, which are incomparable, and capture the remaining vertices through dynamic programming.
\im The second difference is that the tree $T$ is no longer a union of disjoint maximal branches. However, we still want to define a suitable ordering on the endpoints of the relevant edges, so that we can define a similar notion of partial precedence and use \Cref{lem:spanning-tree}. Intuitively, we want a set of disjoint branches, one containing each minimal vertex, such that a variant of \Cref{lem:spanning-tree} still holds, so that we can set up a similar minimum $s$--$t$ cut problem. We handle this issue with \emph{heavy-light decomposition}, a well-known routine that breaks up a tree into long chains, combined with color-coding as before, since we do not know beforehand which chains are useful for us.
\EE

\subsubsection{General Case: Algorithm}\label{sec:gen-alg}

We will perform dynamic programming on the tree $T$, rooted at an arbitrary vertex $r_0$.\footnote{In this section, we will free up variable $r$ to be used as an integer, to be more consistent with the variable choice in Section~2.3.2 of \cite{GLL18b}.} We define the dynamic programming states as follows:
 
\BD[DP State]
For vertex $x\in T$ and integer $k'\in[0,k-1]$, define $\st(x,k')$ as the minimum number of edges cut in $G[V(T(x))]$ over all partitions of $V(T(x))$ obtained by cutting $k'-1$ edges from $T(x)$. More formally,
\begin{gather} \st(x,k') := \min_{S_1,\lds,S_{k'}}|E_{G[V(T(x))]}[S_1,\lds,S_{k'}], \label{eq:st}\end{gather}
where the minimum is over all partitions $S_1,\lds,S_{k'}$ of $V(T(x))$ satisfying $|E_{T(x)}[S_1,\lds,S_{k'}]|=k'-1$. If there is no valid partition $S_1,\lds,S_{k'}$, then $\st(x,k')=\infty$.
\ED
\BO
$|OPT| = \st(r,k-1)$.
\EO
For the rest of this section, we will only be concerned with computing the actual \emph{value} $\st(r,k-1)=|OPT|$. The $k$-cut that achieves this value can be recovered from the dynamic program using standard backtracking procedures (at no asymptotic increase in running time).

The base cases are:
\BE
\im $\st(x,0)=0$ for all vertices $x\in V(T)$, and
\im $\st(x,k')=\infty$ for all leaves $x$ and integer $k'\in[1,k-1]$. 
\EE

Fix a non-leaf vertex $x$, and assume that the values $\st(v,s)$ have already been computed for all $v\in T(x)\setminus\{x\}$ \footnote{For the rest of this section, the reader may assume that $T=T(x)$ for convenience. Every time we refer to $T(v)$, we always have $v\in T(x)$, so $T(v) = (T(x))(v)$, but the latter is more cumbersome to write.} and integer $k'\in[0,k-1]$. We seek to compute the states $\st(x,k')$ for $k'\in[0,k-1]$.

We can easily detect whether or not $\st(x,k')=\infty$: it is $\infty$ iff there are less than $k'-1$ edges in $T(x)$. Therefore, let us also assume that $\st(x,k')<\infty$. Consider the components that achieve the minimum in (\ref{eq:st}), as well as the $k'-1$ edges in $T(x)$ cut by those components. Let the children of these $k'-1$ edges be $v^*_1,\lds,v^*_{k'-1}$. Furthermore, suppose that they are ordered so that $\{v^*_1,\lds,v^*_{k'-1}\}{\downarrow}=\{v^*_1,\lds,v^*_r\}$ for some $r\in[k'-1]$. For each $i\in[r]$, let $k^*_i:=|\{v^*_1,\lds,v^*_{k'-1}\}\cap V(T(v^*_i))| \in [k'-1]$ be the number of these vertices preceded by $v^*_i$, so that $\sum_{i=1}^rk^*_i=k'-1$. 

\BD[HLD]
A \emph{heavy-light decomposition (HLD)} of a tree $T$ is a partition $\m B$ of the edges of $T$ into disjoint branches, such that for each vertex $v\in V(T)$, the branch from $v$ to the root $x$ of $T$ shares edges with at most $O(\logn)$ branches in $\m B$.
\ED

\BF
For any tree, a HLD of the tree exists and can be computed in linear time.
\EF

Fix a HLD of $T(x)$ with branches $\m B$. For a vertex $v\in V(T(x))\setminus\{x\}$, define $B(v)$ to be the branch in $\m B$ containing the parent edge of $v$. For convenience, define $B^*_i:=B(v^*_i)$ for $i\in[r]$; here, $r$ is now an integer, not the root of $T(x)$, since $x$ is now that root. Note that since $v^*_1,\lds,v^*_r$ are pairwise incomparable, the branches $B^*_1,\lds,B^*_r$ are always distinct. For each branch $B\in\m B$, define $\subroot(B)$ to be the child of the root of $B$ that lies on $B$ (hence \emph{sub}root). We now focus on the ``minimal'' branches $B^*_i$, formalized as follows: Let $q:=|\{\subroot(B^*_i):i\in[r]\}{\downarrow}|$, and order $v^*_1,\lds,v^*_r$ so that $\{\subroot(B^*_i):i\in[q]\} = \{\subroot(B^*_i):i\in[r]\}{\downarrow}$. In particular, the vertices $\subroot(B^*_i)$ for $i\in[q]$ are incomparable. Finally, for each $i\in[r]$, define $u^*_i:=\subroot(B^*_i)$.

This time, define $E'\s E$ as the edges whose endpoints $u,v$ satisfy the property that $\subroot(B(u))$ and $\subroot(B(v))$ are incomparable. Define $H$ to be the graph on $\{v^*_1,\lds,v^*_r\}$ with edge set as follows: for each edge $e\in E'$ with endpoints in $T(u^*_i)$ and $T(u^*_j)$ ($i\ne j$, $i,j\in[r]$), add that same edge $e$ between $v^*_i$ and $v^*_j$. The main difference of $H$ compared to \Cref{sec:single} is that this $H$ does not include edges between $v^*_i$ and $v^*_j$ in a common minimal branch $B^*_h$, $h\in[q]$.

For each $i\in[r]$ and vertex $v\in T(v^*_i)$, let $b(v)$ be the first vertex in $B^*_i$ on the branch from $v$ to $r$. Clearly, such a vertex always exists and is inside $T(v^*_i)$. For a vertex set $U\s \bigcup_iT(v^*_i)$, define $b(U):=\bigcup_{u\in U}b(u)$. We now have a lemma resembling \Cref{lem:spanning-tree} from \Cref{sec:single}:

\BL\label{lem:spanning-tree2}
For each connected component $C$ in $H$, there exists a spanning tree $T_C$ of $C$ satisfying the following property: Let $U$ be the set of endpoints of edges in $T_C$. Then, every edge in $C$ is partially preceded by some vertex in $b(U)$.
\EL
\BP
The proof essentially follows from applying \Cref{lem:spanning-tree} on an appropriately chosen graph. Construct the graph $G_b$ as follows: the vertex set is $\bigcup_iB^*_i$, and for each edge in $C$ with endpoints $u,v$ in $G$, add that edge with endpoints $b(u),b(v)$ in $G_b$. Note that if we contract each vertex set $V(B^*_i)$ in $G_b$, we get exactly $H$. Define a tree $T_b$ by taking the branches $B^*_1,\lds,B^*_r$ and adding a root $r_b$ of $T_b$ connected to the root of each $B^*_i$. Apply \Cref{lem:spanning-tree} on the graph $G_b\cup T_b$ (we take union with $T_b$ only to include the root $r_b$), which gives us a spanning tree $T_C$. Let $U$ be the endpoints of $T_C$ in $G$, which means that the set of endpoints in $G_b$ is exactly $b(U)$. By the guarantee of \Cref{lem:spanning-tree} and the definition of partially precede \Cref{def:precede}, every edge $e$ in $C$ with endpoints $u,v$ in $G$ has one of its endpoints $b(u),b(v)$ preceded by a vertex in $b(U)$ in $T_b$, and hence also in $T(x)$. Since $b(u)$ always precedes $u$ in $T(x)$ and the same for $b(v)$ and $v$, one of $u,v$ is preceded by a vertex in $b(U)$ in $T(x)$.
\EP
As in \Cref{sec:single}, let the connected components of $H$ be $C^*_1,\lds,C^*_z$ for $z\le V(H)=r$, ordered in an arbitrary order, and let $T^*_i$ be the spanning tree for $C^*_i$ promised by \Cref{lem:spanning-tree}. We first color the edges of $E'$ identically to \Cref{sec:single-alg}:

\BCO[Same as \Cref{cond:color-i}]\label{cond:color-i2}
For each $C^*_i$, all edges in $T^*_i$ are colored \green.
\ECO
\BCO[Same as \Cref{cond:color-0}]\label{cond:color-02}
All edges in $\bigcup_{i\in[q]}\pt_G(T(u^*_i))\setminus \bigcup_{i}T^*_i$ are colored \red.
\ECO
\BAL[Same as \Cref{alg:color}]\label{alg:color2}
Color each edge in $E'$ \red with probability $1-1/\la$ and \green with probability $1/\la$. With probability $2^{-O(k)}\la^{-k}$, \Cref{cond:color-i2,cond:color-02} hold.
\EAL

\begin{figure}\centering
\begin{tikzpicture}[scale=.5]

\tikzstyle{every node} = [scale=.8,fill,circle]

\draw [magenta, line width=2.5] (0,4) node (v1) {} -- (5,-1) -- (3,1) -- (-1,-4);
\draw[line width=2.5] (10,1) -- (v1.center);
\draw[magenta, line width=2.5] (v1.center) -- (-11,-1);
\draw[  line width=2.5] (-7.4271,0.6424) -- (-11.685,-7.8735);
\draw[gray!70,  line width=2.5] (-9.362,-3.3825) -- (-3.7271,-9.4623);
\draw[ line width=2.5] (-6.6081,-6.4118) -- (-11.1838,-13.1059);
\draw[ line width=2.5] (-4.9345,-8.1701) -- (-6.3539,-11.1146) ;
\draw[purple, line width=2.5] (-6.3539,-11.1146)-- (-5.0193,-14.2287);
\draw[purple,  line width=2.5] (-9.6162,-11.0511) -- (-6.9894,-14.61);
\draw[purple,  line width=2.5] (-8.0486,-13.1271) -- (-9.6162,-16.6013);
\draw[  line width=2.5] (0.5521,-2.1115) -- (5.954,-12.1103);
\draw[purple,  line width=2.5] (2.4586,-5.6916) -- (-0.8461,-11.0934);
\draw[magenta,  line width=2.5]  (4.0686,-8.6785) -- (1.2935,-14.7159);
\draw[ purple,line width=2.5] (8.3689,1.4686) -- (5.8269,-4.9501);
\draw[ purple, line width=2.5] (6.7801,-2.5563) -- (10.0001,-8.742);
\node at (-8.1529,-0.8865) {};
\node at (-10.6042,-5.751) {};
\node at (-7.4162,-7.618) {};
\node at (-9.6643,-10.933) {};
\node at (-9.2705,-15.8483) {};
\node at (-6.3239,-11.0981) {};
\node at (1.2586,-3.4267) {};
\node at (2.4652,-5.6875) {};
\node at (9.3153,-7.4922) {};
\node at (8.3218,1.4815) {};
\node at (1.5197,3.4996) {};

\tikzstyle{every node} = []
\node[above,scale=1.5] at (v1) {$x$};
\node[left,scale=1.5] at (-8.1529,-0.8865) {$u^*_1$};
\node[left,scale=1.5] at (-10.6042,-5.751) {$v^*_1$};
\node[left,scale=1.5] at (-7.4162,-7.618) {$v^*_4$};
\node[right,scale=1.5] at (-7.4162,-7.618) {$u^*_4$};
\node[left,scale=1.5] at (-9.6643,-10.933) {$b(v)$};
\node[left,scale=1.5] (v2) at (-9.2705,-15.8483) {$v$};
\node[left,scale=1.5] at (-6.3239,-11.0981) {$v^*_5$};
\node[right,scale=1.5] at (1.2586,-3.4267) {$u^*_2$};
\node[left,scale=1.5] at (2.4652,-5.6875) {$v^*_2$};
\node[right,scale=1.5] (v3) at (9.3153,-7.4922) {$v'$};
\node[above,scale=1.5] at (8.3218,1.4815) {$b(v')$};
\node[above,scale=1.5] at (1.5197,3.4996) {$u^*_3$};

\draw [green,line width=2](-5.542,-12.9052) -- (-0.0653,-9.7649) node[pos=.4,above,scale=1.5] {$C^*_1$};
\draw [red,line width=2](-5.9942,-11.7998) -- (1.7184,-13.7091) ;

\draw [green,line width=2] plot[smooth, tension=.7] coordinates {(-9.2128,-15.8271) (-.7518,-16.8863) (6.4844,-14.1536) (9.3231,-7.4595)} ;
\node[scale=1.5,green] at (1.9299,-15.7636) {$C^*_2$};
\end{tikzpicture}
\caption{
An example with $r=6$ and $q=3$. There are two components $C^*_1,C^*_2$, and one component $C^+_1$.
 For \Cref{cond:branch1} to hold, none of the  black branches can be contracted. For \Cref{cond:branch2} to hold, the purple branches must be contracted. For \Cref{cond:branch3} to hold, the magenta branches must be contracted. Whether the gray branch is contracted or not does not affect the three conditions.}
\label{f3}
\end{figure}
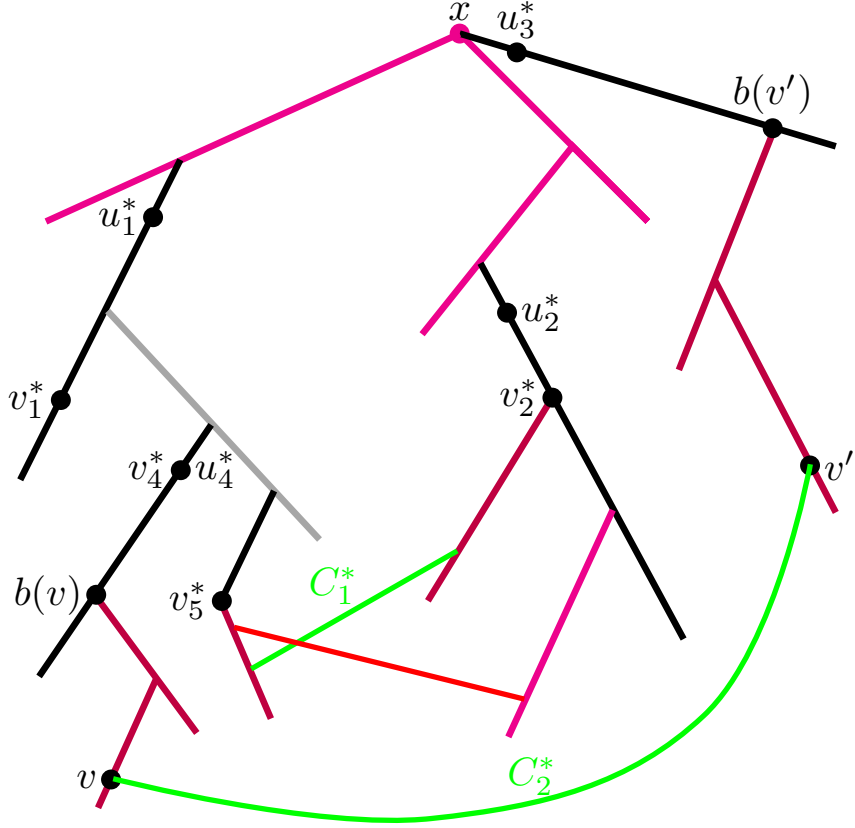

Unlike \Cref{sec:single-alg}, we do not build the graph on the children of the root yet. This time, the algorithm now contracts every branch of the HLD independently with probability $1/\log n$. We would like the following three conditions to hold (see Figure~\ref{f3}):

\BCO\label{cond:branch1}
For each $i\in[r]$, $B^*_i$ is not contracted.
\ECO
\BCO\label{cond:branch2}
For each $C^*_i$ and each edge $e$ in $T^*_i$ with endpoints $u,v$ in $G$, all branches in $\m B$ intersecting the branch from $u$ to $b(u)$ are contracted, and the same holds for the branch from $v$ to $b(v)$.
\ECO
\BCO\label{cond:branch3}
For all $i\in[q]$, all branches in $\m B\setminus \{B^*_i\}$ intersecting the branch from $v^*_i$ to $x$ are contracted.
\ECO

Clearly, \Cref{cond:branch1} holds with probability exactly $1/(\logn)^r$, which is negligible (see \Cref{lognk}). We now claim the following for \Cref{cond:branch2}:
\BCL
Conditioning on the event that \Cref{cond:branch1} holds, \Cref{cond:branch2} is true with probability $2^{-O(r)}$.
\ECL
\BP
Every $T^*_i$ has $O(|V(C)|)$ endpoints in $G$, and for each such endpoint $v$, at most $O(\logn)$ branches in $\m B$ intersect the branch from $v$ to $b(v)$ by the guarantee of HLD. Over all $T^*_i$, this is $\sum_CO(|V(C)|)\cd O(\logn) = O(r\log n)$ many branches. We now show that none of these branches in $\m B$ can be $B^*_j$ for some $j\in[r]$.

Fix $i\in[r]$, and consider an endpoint $v\in T^*_i$. First, suppose for contradiction that the branch from $v$ to $b(v)$ intersects the branch $B^*_i$. Let $(u',v')$ be an edge in $B^*_i$ on the branch from $v$ to $b(v)$, with $v'$ as the child of $u'$. Since $v'$ lies on the branch from $v$ to $b(v)$, we have that $b(v)$ precedes $v'$. But then $v'$ is a vertex on $B^*_i$ encountered before $b(v)$ on the branch from $v$ to $x$, contradicting the choice of $b(v)$.

Next, suppose for contradiction that the branch from $v$ to $b(v)$ intersects the branch $B^*_j$ for $i\ne j$. Extend the branch from $v$ to $b(v)$ into the branch from $v$ to $v^*_i$, which by assumption contains a vertex in $B^*_j$. First, if $v^*_i$ is on $B^*_j$, then both $v^*_i$ and $v^*_j$ lie on a common branch $B^*_j$, contradicting the fact that they are incomparable. Otherwise, the path from $v$ to $v^*_i$ must travel beyond $B^*_j$, so every vertex in $B^*_j$ is preceded by $v^*_i$. In particular, $v^*_j\in V(B^*_j)$ is preceded by $v^*_i$, again contradicting the fact that they are incomparable. 

Therefore, even if we condition on \Cref{cond:branch1}, none of the $O(r\log n)$ relevant branches are automatically contracted (so that the probability of success is not automatically $0$). Thus, the probability that none of the $O(r \log n)$ branches are contracted is $(1-1/\logn)^{O(r\logn)}=2^{-O(r)}$.
\EP


Finally, conditioned on \Cref{cond:branch1,cond:branch2}, \Cref{cond:branch3} holds with probability at least $(1-1/\logn)^{O(r\logn)}=2^{-O(r)}$, since by the HLD property, there are $O(\logn)$ many branches that still need to be contracted for each $v^*_i$ in \Cref{cond:branch3}. Thus, we have the following:

\BAL\label{alg:hld-contr}
Compute a HLD of $T(x)$ into branches $\m B$, and contract each branch with probability $1/\logn$. With probability at least $1/(\logn)^k\cd2^{-O(k)}$, \Cref{cond:branch1,cond:branch2,cond:branch3} hold. Let $T'$ be the resulting tree; note that $x$ is still the root of $T'$. 
\EAL

Define $T'$  as in \Cref{alg:hld-contr}; we have the observation below:
\BO
Assuming \Cref{cond:branch3}, all minimal branches $B^*_1,\lds B^*_q$ now have $x$ as their root.
\EO
Moreover, the following observation follows immediately from \Cref{lem:spanning-tree2}:
\BO
Assume \Cref{cond:branch2}.
Fix any component $C^*_i$, and let $U$ be the set of endpoints in $T'$ of edges in $T^*_i$. Then, every edge in $C^*_i$ is partially preceded by some vertex in $U$ (in the tree $T'$).
\EO


We now define the set $\m U$ similarly to \Cref{sec:mincut-setup}, except this time using the tree $T'$. Build a graph whose vertices are the children of $x$ in $T'$, and for every two children $u,v$ of $x$, connect them by an edge if there is a \green edge between $T'(u)$ and $T'(v)$. For each (maximal) connected component in this graph, add its set $U$ of vertices into a collection $\m U$. 

Next, define the following graph $H'$. Its vertex set is $\{v^*_1,\lds,v^*_q\}$, and it is obtained from $H$ as follows: for each $i\in[q]$, contract into $v^*_i$ all vertices $v^*_j$ ($j\in[r]$) such that $u^*_j$ is preceded by $u^*_i$. For each connected component $C$ of $H'$, let $C^+$ be the set of vertices in $H$ contracted to a vertex in $C$. Every $C^+$ is a union of some connected components $C^*_i$ of $H$; 
let $C^+_1,\lds,C^+_y$ be all such vertex sets $C^+$. 

\BL
If \Cref{cond:color-02,cond:color-i2,cond:branch3} hold, then for each $C^+_i$, there exists a set $U^+_i\in\m U$ whose subtrees contain precisely all vertices $v^*_i$ in $C^+_i$ (and no more).
\EL
\BP
Consider a vertex set $C^+_i$ whose vertices get contracted into $v^*_{i_1},\lds,v^*_{i_\el}$ in $H'$. For each vertex $v^*_h$ in $C^+_i$, there is a vertex $v^*_{i_j}$ ($i\in[\el]$) such that $u^*_h$ is preceded by $u^*_{i_j}$, which means that $v^*_h$ is also preceded by $u^*_{i_j}$. Therefore, all vertices $v^*_h$ in $C^+_i$ must be inside $T'(u^*_{i_j})$ for some vertex $v^*_{i_j}$.

By \Cref{cond:branch3}, all the vertices $u^*_{i_1},\lds,u^*_{i_\el}$ are children of the root $x$ of $T'$. By \Cref{cond:color-i2}, there must be \green edges connecting each connected component $C^*_h\s C^+$. These components $C^*_h$ are connected together through vertices in different $C^*_h$ getting contracted into the same $v^*_{i_j}$, which means they share vertices in a common $T'(u^*_{i_j})$. It follows that all vertices in $C^+_i$ belong to subtrees in a single $U\in\m U$. By \Cref{cond:color-02}, there cannot be any \green edges in $\bigcup_{i_j}\pt_G(T(u^*_{i_j}))\setminus\bigcup_iT^*_i$, so there cannot be any other vertices $v^*_h$ in $U$.
\EP

\BA\label{as:good2}
After running \Cref{alg:color2,alg:hld-contr}, Conditions~\ref{cond:color-i2},~\ref{cond:color-02},~\ref{cond:branch1},~\ref{cond:branch2},~and~\ref{cond:branch3} hold, and for each $C^+_i$, there exists a set $U^+_i\in\m U$ whose subtrees contain precisely all vertices $v^*_j$ in $C^+_i$ (and no vertices $v^*_j$ outside $C^+_i$). (Note that $\bigcup_iU^+_i=\{u^*_1,\lds, u^*_q\}$. This assumption holds with probability at least $2^{-O(k)}(\logn)^{-k}\la^{-k}$.)
\EA

Before we move on to the next section, for each $U\in\m U$,  let us define
\[\minelts(U):=\big( \bigcup_{u\in U} \{v : v\in T'(u)\text{ and } \exists(v,v')\in\pt T'(u)\text{ \green}\}\big){\downarrow}.\]
Note that if we assume \Cref{cond:color-i2,cond:color-02,cond:branch2}, then for each $ U^+_i=\{u^*_{i_1},\lds,u^*_{i_\el}\}$, we also have \linebreak$\minelts(U^+_i) =\big( \bigcup_{j} \{b(v) : v\in T(u_{i_j})\text{ and } \exists(v,v')\in\pt T(u_{i_j})\text{ \green}\}\big){\downarrow} \s\bigcup_jB^*_{i_j}$.  After constructing $\minelts(U)$ for each $U\in\m U$, the algorithm may forget the construction of $T'$; indeed, $T'$ was only needed to construct the sets $U$ and $\minelts(U)$. However, we will still need $T'$ for our analysis.

\subsubsection{General Case: Processing the Sets $U\in\m U$}\label{sec:gen-cut}

We begin similarly to \Cref{sec:mincut-setup}: define a similar \emph{minimum ancestor $p$-cut} problem for $p\in[k'-1]$:
\BD[Minimum ancestor $p$-cut]
Fix some set $U=\{u_1,\lds,u_\el\}\in\m U$, and let $\minelts(U)=\{s_1,\lds,s_m\}$. The \emph{minimum ancestor $p$-cut} is the following problem: For each $i\in[\el]$, remove at least one edge in $T(u_i)$ (note: not $T'(u_i)$) such that after removal, no $s_j\in V(T(u_i))$ is in the same component as $u_i$, and such that exactly $p-1$ edges are removed in total (over all $i$). Consider the $p$-cut in $G$ from the connected components of the remaining forest. We want to compute the $p$-cut of minimum size, denoted $\minanccut(U,p)$.
\ED

\BCL
Assuming \Cref{as:good2}, for each $ U^+_i=\{u^*_{i_1},\lds,u^*_{i_\el}\}$ with $p^+_i:=|\bigcup_jV(T(u^*_{i_j}))\cap\{v^*_1,\lds,v^*_{k'-1}\}|$, we have $|\pt_G[S^*_{i_1},\lds,S^*_{i_\el}]|=|\minanccut( U^+_i,p^+_i)|$. Moreover, $\sum_{U^+_i}|\minanccut(U^+_i,p^+_i)|=OPT$.
\ECL
\BP
The proof resembles the proof of \Cref{clm:anc}, except with the additional HLD and \Cref{as:sqrt} added in.

Let the set $C^+_i$ have vertices $v^*_{i_1},\lds,v^*_{i_{\el'}}$ for $\el'\ge\el$. Assume that the components $C^*_1,\lds,C^*_y$ are reordered so that $C^*_1,\lds,C^*_z \s C^+_i$ are the connected components of $H$ whose vertices form $C^+_i$. For simpler notation, define $p:=p^+_i$, and let $v^*_{i_{\el+1}},\lds,v^*_{i_{p}}$ be the other vertices in $\bigcup_jV(T'(u^*_{i_j}))$, that is, those in $\bigcup_jV(T'(u^*_{i_j}))\cap\{v^*_{r+1},\lds,v^*_{k'-1}\}$. First, we claim that a $p$-ancestor cut of size $|\pt_G[S^*_{i_1},\lds,S^*_{i_\el}]|$ is achievable: simply cut the parent edges of $v^*_{i_1},\lds,v^*_{i_p}$. We now show that this cut is indeed an ancestor cut. By \Cref{cond:color-i2,cond:color-02}, the only \green edges in any $\pt_G(T(u^*_{i_j}))$ lie in some $T^*_i$, $i\in[z]$. By \Cref{cond:branch2}, all endpoints of $T^*_1,\lds,T^*_z$ in $T'$ are on the branches $B^*_{i_1},\lds,B^*_{i_{\el'}}$. Inside each branch $B^*_{i_j}$, vertex $v^*_{i_j}$ precedes any endpoint in $T(x)$ of a \green edge on $B^*_{i_j}$, and these branches cover all \green edges. Therefore, every vertex in $\minelts(U^+_i)$ is preceded by some vertex $v^*_{i_j}$ in $T'$. Since $T'$ is a contraction of $T(x)$, precedence is unchanged, so every vertex in $\minelts(U^+_i)$ is also preceded by some vertex $v^*_{i_j}$ in $T(x)$. In other words, the parent edges of $v^*_{i_j}$ lie on the branches in $T(x)$ from vertices in $\minelts(U^+_i)$ to $x$,  so our cut is a valid ancestor cut.

The proof that no better ancestor cut is possible is identical (see the proof of \Cref{clm:anc}), so we omit it.
\EP

We now construct our graph $G'$, similar to the one in \Cref{sec:mincut-setup}. Fix a set $U=\{u_1,\lds,u_\el\}$ and let $\minelts(U)=\{s_1,\lds,s_\el\}$ where $s_i\in V(T(u_i))$. For each $p\in[k'-1]$, we will compute a value $f(U,p)$. First, we construct the following multigraph $G'$ on the vertices $\lp\bigcup_i V(T(u_i))\rp\cup \{x\}$:

\BE 
\im[(a)] For each edge in $E\setminus E'$ with both endpoints in $T(\minelts(U))$, do nothing: these edges are not cut in any ancestor cut.
\im[(b)] For each edge in $E'$ with at least one endpoint in $T(\minelts(U))$, do nothing: these edges are \emph{always} cut in an ancestor cut, and we will account for these edges separately. 
\im[(c)] For each edge in $E\setminus E'$ with at most one endpoint in $T(\minelts(U))$, add it to $G'$.
\im[(d)] For each edge $(u,v)\in E'$ with both endpoints not in $T(\minelts(U))$, add edges $(u,r)$ and $(v,r)$ to $G'$.
\EE
Note that all edges in $G'$ have both their endpoints in $V(T(u_i))\cup \{x\}$ for some $i$. Next, for each sequence $p_1,\lds,p_\el$ of positive integers summing to $p$, do the following: For each $i\in[\el]$, consider all ways to select $p_i$ vertices $v_1,\lds,v_{p_i}\in T(u_i)$ so that if we remove their parent edges, then no vertex in $\minelts(U)\cap V(T(u_i))$ is in the same component as $u_i$; find the way that minimizes $|\bigcup_j\pt_{G'}(T(v_j))|$. Then, sum over the costs of the $(p_i+1)$-cuts for each $i\in[\el]$. Finally, compute the minimum sum over all sequences $p_1,\lds,p_\el$, and add the number of edges considered in step (b) to this minimum. The final value is $f(U,p)$.

It is not clear how to compute $f(U,p)$ quickly, and this is where we will use the previously computed $\st(\cd,\cd)$'s; we defer this to \Cref{sec:computing}. 
The intuition for the construction of $G'$ is the same as in \Cref{sec:mincut-setup}.

Again, define the following natural correspondence between ancestor cuts and the ``$G'$-cuts'': two correspond to each other if their cut edges are identical.
\BL\label{lem:upper-bound2}
For two corresponding cuts, the size of the ancestor cut is at most the size of the $G'$-cut plus the number of edges in step~(b).
\EL
\BP
The proof is identical to the one in \Cref{lem:upper-bound}, so we omit it.
\EP
\BL\label{lem:equality2}
Assuming \Cref{as:good}, for each $U^+_i$ and $p^+_i$,  $\minanccut(U^+_i,p^+_i)$ has size exactly $f(U^+_i,p^+_i)$.
\EL
\BP
In the proof of \Cref{lem:upper-bound} (adapted to suit \Cref{lem:upper-bound2}), the only potential source of inequality is in (d): an edge with no endpoints in $x$'s component of $\minanccut(U^+_i,p^+_i)$ contributes $1$ to $|\minanccut(U^+_i,p^+_i)|$ and $2$ to the corresponding $G'$-cut. If such an edge $e$ existed, then it must be in $\pt_G[S^*_{i_1},\lds,S^*_{i_\el}]$ where $V(C^+_i):=\{v^*_{i_1},\lds,v^*_{i_\el}\}$, which means $e$ is in component $C^+_i$. Also, neither of its endpoints is preceded by a vertex in $\minelts(U^+_i)$, which means that for any edge in any $C^*_j\s C^+_i$ with endpoints $u$ and $v$ in $G$,  neither of $e$'s endpoints is preceded by $b(u)$ or $b(v)$. In particular, this is true for any edge in any $T^*_j$ as well. We thus have an edge $e$ in some $C^*_j\s C^+_i$ not partially preceded by any endpoint in $T^*_j$, contradicting the definition of $T^*_j$. Therefore, no such edge exists, and we have equality.\EP

Let us first assume that for each $U\in \m U$ and $p\in[k'-1]$, we can compute $f(U,p)$ in time $n^{o(k)}$. (Unlike the one in \Cref{sec:mincut-setup}, this assumption is nontrivial and is covered in \Cref{sec:computing,sec:sqrt}.) Then, as in \Cref{sec:mincut-setup}, we can compute the minimum similar to (\ref{eq:sum-f}) as a knapsack problem:

\BAL\label{alg:knapsack2}
Compute 
\begin{gather} \min_{\substack{U_1,\lds,U_\el \\ \sum_i|U_i|=r-1 \\ p_1,\lds,p_\el \\ \sum_i p_i = k'-1}} \sum_{i=1}^\el f(U_i,p_i) . \label{eq:sum-f2} \end{gather}
 in polynomial time by formulating it as a knapsack problem. Assuming \Cref{as:good2}, the result is exactly $OPT$.
\EAL

We have the final algorithm below, which proves \Cref{thm:tree}:

\BAL\label{alg:final-single}
For $O(2^{O(k)}\la^k\logn)$ repetitions, run \Cref{alg:color2,alg:knapsack2}, and output the minimum value of (\ref{eq:sum-f2}) ever computed.
\EAL

\subsubsection{Computing $f(U,p)$}\label{sec:computing}

In this section, we describe how to compute $f(U,p)$ in a rather ad-hoc way, which is where we pick up the $n^{o(k)}$ multiplicative factor. Note that it can be easily computed in $O(k^{O(k)}n^{p+O(1)})$ time by brute force, which is fine if $p=O(1)$ or even $o(k)$. However, $p$ could be as large as $\Th(k)$.

First, let us make an additional assumption for now. We later show how to remove this assumption, at a cost of a multiplicative $n^{o(k)}$ factor.
\BA\label{as:sqrt}
For each  $U^+_i=\{u^*_{i_1},\lds,u^*_{i_\el}\}$, we have $|V(T'(u^*_{i_j}))\cap\{v^*_1,\lds,v^*_r\}|\le\sr k$ for all $j\in[\el]$. (Note that $V(T'(u^*_{i_j}))\cap\{v^*_1,\lds,v^*_r\}$ is also $\big(V(T'(u^*_{i_j}))\cap\{v^*_1,\lds,v^*_{k'-1}\}\big){\downarrow}$.)
\EA

In this case, the algorithm computes an estimate $f'(U,p)$ of $f(U,p)$ so that $f'(U,p)\ge f(U,p)$ always, and $f'(U^*_i,p)=f(U^*_i,p)$. First, fix a set $U=\{u_1,\lds,u_\el\}\in\m U$ and integer $p\in[k'-1]$. Next, for each sequence $p_1,\lds,p_\el$ of positive integers summing to $p$, for each sequence of positive integers $r_1,\lds,r_\el$ with $r_i\le\min\{p_i,\sr k\}$ (for all $i\in[\el]$), do the following: For each $i\in[\el]$, consider all ways to select $r_i$ incomparable vertices $v_1,\lds,v_{r_i}\in T'(u_i)\setminus \{u_i\}$ and remove their parent edges so that no vertex in $\minelts(U)\cap V(T(u_i))$ is in the same component as $u_i$; find the way that minimizes
\begin{gather} \min_{\substack{k'_1,\lds,k'_{r_i}\ge1\\\sum_jk'_j=p_i}} \sum_{j=1}^{r_i}\st(v_j,k'_j) + \bigg| \bigcup_{j=1}^{r_i} \pt_{G'}(T(v_j)) \bigg| 
.\label{eq:mink}\end{gather}
Let the minimum of (\ref{eq:mink}) over all selections of $v_1,\lds,v_{r_i}$ be $M_i$. Finally, compute the minimum sum $\sum_{i\in[\el]}M_i$ over all sequences $p_1,\lds,p_\el$, and add the number of edges considered in step (b) to this minimum. The final value is $f'(U,p)$.
 Clearly, $f'(U,p)$ can be computed in $O(k^{O(k)}n^{\sr k+O(1)})$ time.

\BCL\label{clm:upper}
For all $U\in\m U$ and $p\in[k'-1]$, $f'(U,p)\ge f(U,p)$.
\ECL
\BP
Let $U=\{u_1,\lds,u_\el\}$ as before. Given $i\in[\el]$ and values $p_i,r_i$ and vertices $v_1,\lds,v_{r_i}$, (\ref{eq:mink}) represents the way to select $(p_i-r_i)$ vertices $v_{r_i+1},v_{r_i+2},\lds,v_{p_i}$ so that $v_1,\lds,v_{p_i}$ form an ancestor $p_i$-cut for $U$ and $\{v_1,\lds,v_{p_i}\}{\downarrow}=\{v_1,\lds,v_{r_i}\}$. This is because the values $\st(v_j,k'_j)$ capture the way to cut $k'_j-1$ additional edges inside each $T(v_j)$ that minimizes the number of edges cut in $G_j$. Note that which edges we cut in each $T(v_j)$ does not affect whether $v_1,\lds,v_{p_i}$ is a valid ancestor cut, since the component containing $u_i$ in the $G'$-cut is already determined by $\{v_1,\lds,v_{p_i}\}{\downarrow}=\{v_1,\lds,v_{r_i}\}$.

Therefore, $f'(U,p)$ is essentially computing the same as $f(U,p)$, except that in $f'(U,p)$, the values $r_i$ are artificially limited by $\sr k$. So $f'(U,p)$ can only be larger than $f(U,p)$.
\EP
\BCL\label{clm:equal}
Assuming \Cref{as:sqrt}, for each $U^+_i$, $f'(U^+_i,p)=f(U^+_i,p)$.
\ECL
\BP
By the proof of \Cref{clm:upper}, the only way for $f'(U^+_i,p)>f(U^+_i,p)$ to happen is if the optimal selection of $r_1,\lds,r_\el$ has $r_i>\sr k$ for some $i$. However, this is precisely what \Cref{as:sqrt} prevents.
\EP

\subsubsection{Removing \Cref{as:sqrt}}\label{sec:sqrt}
In this section, we deal with \Cref{as:sqrt} in another ad-hoc way. Essentially, we show that if \Cref{as:sqrt} does not hold, then we can preprocess the tree $T(x)$ in an efficient way (more precisely, with $n^{o(k)}$ multiplicative overhead) so that \Cref{as:sqrt} does hold.

\BD[Rank]
Given a vertex $u^*_i$, $i\in[q]$ (that is, $u^*_i$ is in some $U^+_{i'}$), define the \emph{rank} of a vertex $v\in T(u^*_i)$ as follows: Consider the branch from $v$ to $T(u^*_i)$, which we call $B$. If $B$ contains any vertex $v^*_j$, then the  $\ra(v)=-\infty$. Otherwise, let $\ra(v)$ be the number of vertices $v^*_j\in T(u^*_i)$ such that if we travel along the branch from $v^*_j$ to $u^*_j$, then we encounter a vertex in $B$ before or at the same time as we encounter a vertex in $\{u^*_1,\lds,u^*_r\}\setminus\{u^*_j\}$.
\ED

\BL\label{rank}
Fix $i\in[q]$, and let $a$ be the maximum rank of a vertex in $T(u^*_i)$. Then,\linebreak $\big|\big(V(T(u^*_{i_j}))\cap\{v^*_1,\lds,v^*_{k'-1}\}\big){\downarrow}\big| \le2^a$.
\EL
\BP
We induct on $a>1$, with the trivial base case $a=0$. Now suppose that $a>0$. Consider the set $S:=\big(\{u^*_1,\lds,u^*_r\}\cap (V(T(u^*_i))\setminus\{u^*_i\}) \big){\downarrow}$.

If $S=\emptyset$, then $v^*_i$ is the only vertex in $\{v^*_1,\lds,v^*_r\}\cap T(u^*_i)$, so $a=1$ and the bound holds. Otherwise, let $S=\{u^*_{i_1},\lds,u^*_{i_\el}\}$; we claim that for all $j\in[\el]$, every vertex in $T(u^*_{i_j})$ has rank at most $a-\el$. Suppose not: there exists $j\in[\el]$ and a vertex $v\in T(u^*_{i_j})$ with rank more than $a-\el$. Let $B'$ be the branch from $v$ to the parent of $u^*_{i_j}$. Then, there are more than $a-\el$ vertices $v^*_{i'}$ such that if we travel the branch from $v^*_{i'}$ to the parent of $u^*_{i_j}$, then we encounter a vertex in $B'$ no later than we encounter some $u^*_{i''}$ for $i''\ne i'$. Extend the branch $B'$ to $B''$ so that $B''$ travels from $v$ to the parent of $u^*_i$; clearly, the previous statement still holds if we travel the branch from $v^*_{i'}$ to the parent of $u^*_i$ instead, and with $B'$ replaced by $B''$. Moreover, for each $j'\ne j$, if we travel the branch from $v^*_{i_{j'}}$ to $u^*_i$, then we also encounter a vertex on $B''$ no later than we encounter any vertex in $\{u^*_1\lds,u^*_r\}\setminus\{u^*_{i_{j'}}\}$ (which has to be $u^*_i$). Thus, these $\el-1$ vertices $v^*_{i_{j'}}$ increase the rank of $v$ in $T(u^*_i)$ to more than $(a-\el)+(\el-1)=a-1$. Finally, $v^*_i$ always increases the count by $1$, so $v$ has rank more than $a$, contradicting the assumption that the maximum rank inside $T(u^*_i)$ is $a$. 

By induction, each $u^*_{i_j}\in S$ satisfies $ \big|\big(V(T'(u^*_{i_j}))\cap\{v^*_1,\lds,v^*_{k'-1}\}\big){\downarrow}\big| \le2^{a-\el}$. Therefore,
\begin{align*}
\big|\big(V(T'(u^*_{i_j}))\cap\{v^*_1,\lds,v^*_{k'-1}\}\big){\downarrow}\big| &=\bigg|\{v_i\} \cup \bigcup_{j=1}^\el \big(V(T(u^*_{i_j}))
\cap\{v^*_1,\lds,v^*_{k'-1}\}\big){\downarrow} \bigg| \\&\le 1 + \el \cd 2^{a-\el} \\&<1+2^\el\cd2^{a-\el}\\&=1+2^a,
\end{align*}
as desired.
\EP

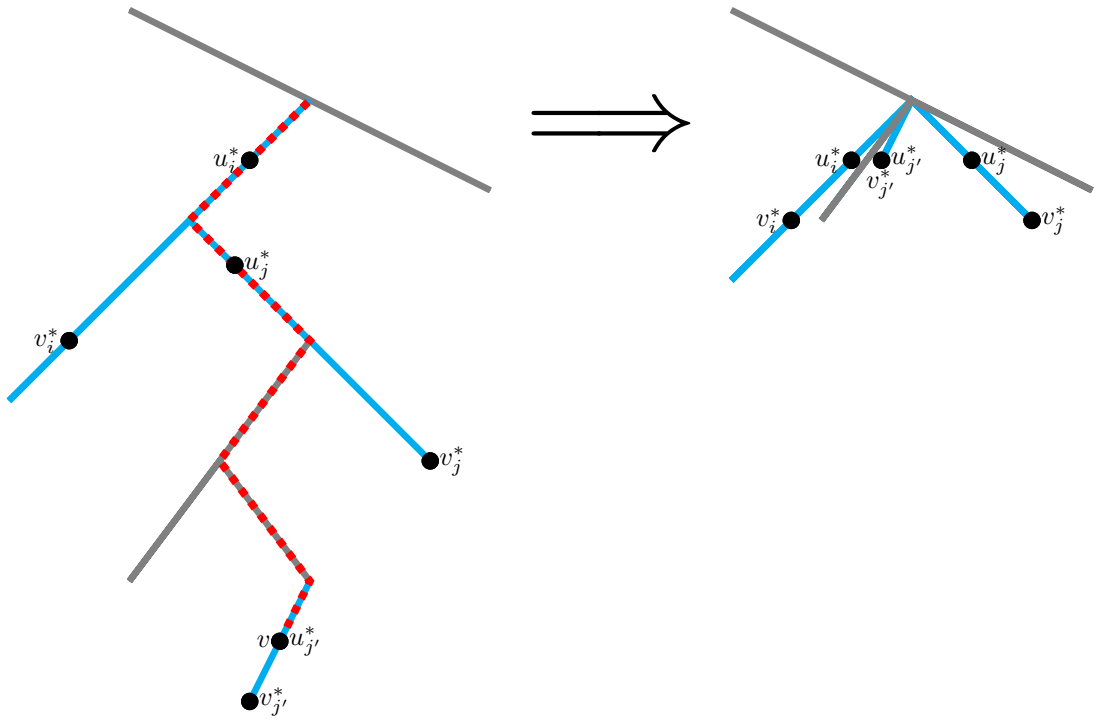
\begin{figure}
\begin{tikzpicture}[scale=.8]

\node at (-1,4) {};
\draw (-6,-1) node (v9) {} -- (-1,4) node (v8) {} -- (-4,5.5) node (v1) {} -- (2,2.5) node (v3) {};
\draw (-3,2) node (v10) {} -- (1,-2) -- (-1,0) node (v4) {} -- (-4,-4) node (v5) {};
\draw (-2.5,-2) {};
\draw (-2.5,-2) node (v6) {} -- (-1,-4) node (v7) {} -- (-2,-6);

\tikzstyle{every node}=[circle, fill, scale=.7];
\node at (-2,3) {};
\node at (-5,0) {};
\node at (-2.2473,1.256) {};
\node (v11) at (1,-2) {};
\node at (-1.5,-5) {};
\node (v12) at (-2,-6) {};
\node at (8,3) {};
\node at (7,2) {};
\node at (10,3) {};
\node at (8.5,3) {};
\node at (11,2) {};
\tikzstyle{every node}=[];

\draw (6,5.5) node (v13) {} -- (12,2.5) node (v14) {};
\draw[cyan,line width=2.5] (9,4) -- (6,1) -- (9,4) node (v2) {} -- (11,2) -- (v2.center) -- (8.5,3) -- (v2.center) -- (7.5,2) node (v15) {};

\draw[gray,line width=2.5] (v1.center) -- (v3.center);
\draw[gray,line width=2.5] (v4.center) -- (v5.center) -- (v6.center) -- (v7.center);
\draw[cyan,line width=2.5] (v8.center) -- (v9.center) -- (v10.center) -- (v11.center);
\draw[cyan,line width=2.5] (v7.center) -- (v12.center);
\draw[gray,line width=2.5] (v13.center) -- (v14.center) -- (v2.center) -- (v15.center) -- (v2.center);

\node[left] at (-2,3) {$u^*_i$};
\node[left] at (-5,0) {$v^*_i$};
\node[right] at (-2.2473,1.256) {$u^*_j$};
\node[right](11) at (1,-2) {$v^*_j$};
\node[right] (v16) at (-1.5,-5) {$u^*_{j'}$};
\node[left] (v166) at (-1.5,-5) {$v$};
\node[right] (12) at (-2,-6) {$v^*_{j'}$};
\node[left] at (8,3) {$u^*_i$};
\node[left] at (7,2) {$v^*_i$};
\node[right] at (10,3) {$u^*_j$};
\node[right] at (8.5,3) {$u^*_{j'}$};
\node[below] at (8.5,3) {$v^*_{j'}$};
\node[right] at (11,2) {$v^*_j$};
\draw[red,line width=3, dashed] (-1.5,-5) -- (v7.center) -- (v6.center) -- (v4.center) -- (v10.center) -- (v8.center);

\tikzstyle{every node}=[circle, fill, scale=.5];
\node at (-2,3) {};
\node at (-5,0) {};
\node at (-2.2473,1.256) {};
\node (vv11) at (1,-2) {};
\node at (-1.5,-5) {};
\node (vv12) at (-2,-6) {};
\node at (8,3) {};
\node at (7,2) {};
\node at (10,3) {};
\node at (8.5,3) {};
\node at (11,2) {};
\tikzstyle{every node}=[];

\node[scale=4] at (4,3.5) {$\Longrightarrow$};

\tikzstyle{every node}=[circle, fill, scale=.7];
\node at (-2,3) {};
\node at (-5,0) {};
\node at (-2.2473,1.256) {};
\node (v11) at (1,-2) {};
\node at (-1.5,-5) {};
\node (v12) at (-2,-6) {};
\node at (8,3) {};
\node at (7,2) {};
\node at (10,3) {};
\node at (8.5,3) {};
\node at (11,2) {};
\tikzstyle{every node}=[];

\end{tikzpicture}
\caption{Contracting the branch from $v$ to the parent of $u^*_i$ (dashed red). The rank of $v$ is $3$. After the contraction, for the vertices $v^*_j$ and $v^*_{j'}$, their new vertices $u^*_j$ and $u^*_{j'}$ are no longer preceded by any other $u^*_{i'}$.}
\label{fs}
\end{figure}

By \Cref{rank}, either \Cref{as:sqrt} holds, or there exists some $i\in[q]$ such that $a\ge \log_2(\sr n)$. In the former case, we compute $f'(U,p)$ as before. In the latter case, consider this integer $i$ and a vertex $v$ with $\rank(v)\ge\log_2(\sr n)$. Contract the branch from $v$ to the parent of $u^*_i$, and consider the new contracted graph with the same HLD (except with some edges contracted). Let the branch from $v$ to $u^*_i$ be $B$. Observe that for each vertex $v^*_j\in V(T(u^*_i))$ satisfying the property in the definition of rank,\footnote{Namely, ``if we travel along the branch from $v^*_j$ to $u^*_j$, then we encounter a vertex in $B$ before or at the same time as we encounter a vertex in $\{u^*_1,\lds,u^*_r\}\setminus\{u^*_j\}$.'' } the vertex $u^*_j$ is no longer preceded by any other $u^*_{j'}$ (see Figure~\ref{fs}). Previously, this property was only true for $v^*_i$. Therefore, if we perform the contraction, keep the contracted HLD, and compute the new values $u^*_1,\lds,u^*_r$, then $q$ increases by (at least) $\ra(v)-1$. We can repeat this process at most $\big\lc \f{k}{\log_2(\sr k)-1} \big\rc$ times before \Cref{as:sqrt} must hold, since otherwise, $q$ would be greater than $k$. Let $t^\dag$ be the number of repetitions, let $v^\dag_1,\lds,v\dag_{t}$ the selected vertices, and let $T^\dag$ be the contracted tree.

Our algorithm is as follows. Start with an arbitrary HLD of $T(x)$. Choose a random number  $t\in\big[\big\lc\f{k}{\log_2(\sr k)-1}\big\rc\big]$, and choose $t$ random vertices $v_1,\lds,v_t\in V(T(x))$. Contract all the edges on the branches from each $v_i$ to $x$. With probability $n^{-o(k)}$, we have $t=t^\dag$ and $v_i=v^\dag_i$ for each $i\in[t]$. Note that in this case, our contracted tree is $T^\dag$ with possibly more edges contracted above any $u^*_i$. Therefore, the new HLD with branches $\m B$ on our new tree satisfies \Cref{as:sqrt}. We then proceed with \Cref{alg:color2,alg:hld-contr}, except we do not recompute the HLD in \Cref{alg:hld-contr}. The algorithms in \Cref{sec:gen-cut,sec:computing} are the same, except we now have \Cref{as:sqrt}. Finally, repeat the entire algorithm $n^{o(k)}$ times, starting from guessing $v$ and $v_1,\lds,v_t$, so that w.h.p., \Cref{as:sqrt} holds in one of the trials.






\section{Kawarabayashi-Thorup Sparsification}\label{sec:KT}

In this section, we prove the following:

\KT*

The algorithm iteratively contracts vertices in $G$, possibly producing a multigraph. It is morally the same as the one in \cite{flow,KT}, except with worse bounds since we are in the approximate, $k$-cut setting. The one key difference is that since we are not concerned with a near-linear running time, we replace their inner PageRank/UnitFlow subroutines with a simpler one that iteratively computes low-conductance cuts. Our algorithm is described in pseudocode in Algorithm~\ref{alg:KT}. Below, we introduce the terminology from \cite{KT} that we use in our algorithm and analysis.

\para{Terminology}
Here, we list the terminology specific to \Cref{sec:KT}.

For a vertex subset $S$, define $\vol(S)=\sum_{v\in S}\deg(v)$ as the sum of degrees of vertices in $S$.
A connected subgraph $H\s G$ is \emph{cut} by edge set $F\s E$ if $H - F$ is disconnected. A set of vertices $C\s V$ is \emph{cut} by $F$ if $G[C]-F$ is disconnected.

The terminology below originate from \cite{KT}. In all the definitions, the graph $\ol G$ is fixed and has $m$ edges.
\BD[Regular vertex and supervertex]
A vertex $v$ in $\ol G$ is a \emph{supervertex} if more than one vertex in $G$ contracts to $v$. Otherwise, it is a \emph{regular vertex}.
\ED

\BO\label{obs:reg}
Every regular vertex $v\in V(\ol G)$ satisfies $\deg_{\ol G}(v)\ge\de$.
\EO

\BD[Passive supervertex]
A vertex $v\in V(\ol G)$ is a \emph{passive supervertex} if it is a supervertex with $\deg_{\ol G}(v)\le 3\al\de/\g$. ($\g:=\f1{100\log m}$ from \Cref{def:expander}.)
\ED

\BD[Conductance]
Given a graph $H$, a set $S:\emptyset\subsetneq S \subsetneq V(H)$ has \emph{conductance} \[ \f {|\pt_HS|} {\min\{\vol(S),\vol(V(H)\setminus S) } .\]
\ED

\BD[$\g$, $\g$-expander]\label{def:expander}
Define $\g:=\f1{100\log m}$ throughout this entire section. A graph $H$ is a \emph{$\g$-expander} if every set $S\s V(H)$ has conductance at most $\g$.
\ED

\BD[Trim]
Let $H$ be a subgraph of $\ol G$. By \emph{trimming} $H$, we mean iteratively removing from $H$ any vertex $v\in V(H)$ satisfying $\deg_H(v)<\f25\deg_{\ol G}(v)$, until no such vertices are left.
\ED

\BD[Loose vertex]
Let $C$ be a subgraph of $\ol G$. A vertex $v\in V(C)$ is \emph{loose} if it is a regular vertex and $d_C(v)\le d_{\ol G}(v)/2$.
\ED

\BD[Shave]
Let $C$ be a subgraph of $\ol G$. By \emph{shaving} $C$, we mean (simultaneously) removing all loose vertices in $C$. (Unlike trimming, this operation is not iterative.)
\ED

\BD[Scrap]
Let $C$ be a subgraph of $\ol G$. Suppose we shaved $C$ into a subgraph $A\s H$. By \emph{scrapping} $A$, we mean removing all vertices from $A$ if $\vol_{\ol G[C]}(A)\le\vol_{\ol G}(C)/4$ (and doing nothing otherwise).
\ED

\BD[Core]
Let $C$ be a subgraph of $\ol G$. The \emph{core} is the subgraph left over after shaving $C$ into $A$ and then scrapping $A$ (that is, replacing $A$ with $\emptyset$ if $\vol_{\ol G[C]}(A)\le\vol_{\ol G}(C)/4$.)
\ED


We first begin with a sparsification algorithm that makes calls to an \emph{exact} minimum conductance cut algorithm. Of course, since conductance is an NP-hard problem, this algorithm is not polynomial-time. Then, we explain at the end how to make the algorithm approximate. We choose this approach for two reasons. First, the algorithm utilizing exact conductance cuts is cleaner to state and analyze, and transitioning to the approximate conductance case is straightforward. Second, for potential future applications which do not require a polynomial time algorithm (and mainly existence of the sparsification), the exact algorithm would suffice and give better bounds, so nothing needs to be reproven.


\begin{algorithm}[H]
\mylabel{KT}{\texttt{KT-Sparsification}}
\caption{\ref{KT}($G=(V,E),\al$)}
\begin{algorithmic}[1]
\State $\ol G\gets G$; $G$ has min degree $\de$
\While{$<\f1{20}$ fraction of the edges in $\ol G$ are incident to passive supervertices}\label{line:2}
  \State $H\gets \ol G$
  \State Remove passive supervertices from $H$ and trim $H$
  \While{there exists a connected component $C$ in $H$ such that $H[C]$ is not a $\g$-expander}
    \State Compute a $\g$-conductance cut in $H[C]$, remove these edges from $H$, and trim $H$\label{line:cut}
  \EndWhile
  \State Take each connected component of $H$ and contract its core (if nonempty) to a supervertex in $\ol G$\label{line:core}
\EndWhile
\State\Return$\ol G$
\end{algorithmic}\label{alg:KT}
\end{algorithm}



For the rest of this section, let us state the assumption on $\de$ in \Cref{thm:KT} as a formal assumption below:

\BA \label{as:k-de}
$\de$ is larger than some absolute constant, and
\begin{gather}
\de > \om(\max\{\al\logn,\al k\}) . 
\nonumber
\end{gather}
\EA

For the rest of this section, we divide the proof of \Cref{thm:KT} into three parts. In \Cref{sec:KT-c}, we show that the graph $\ol G$ returned by Algorithm~\ref{alg:KT} satisfies Condition~1 of \Cref{thm:KT}, the ``correctness'' guarantee. In \Cref{sec:KT-q}, we show that $\ol G$ satisfies Condition~2, the ``quality'' guarantee. Finally, in \Cref{sec:KT-t}, we show that Algorithm~\ref{alg:KT} indeed terminates, rather than looping indefinitely at line~\ref{line:2}.

\subsection{Correctness}\label{sec:KT-c}

In this section, we will prove the following lemma, which argues that the algorithm correctly preserves nontrivial $k$-cuts of small enough size (Condition~1 of \Cref{thm:KT}):
\BL[Condition~1 of \Cref{thm:KT}]\label{lem:correctness}
Suppose that the optimal $k$-cut has size~$\le\al\de$ in $G$. Then, every non-optimal nontrivial $k$-cut is preserved in $\ol G$. That is, no edge of the cut is contracted in $\ol G$.
\EL

The structure of the proof follows that of \cite{KT}, except adapted to the approximate and $k$-cut settings.

\BL\label{lem:S-size-bound}
Consider a point in the algorithm's execution, just after it trimmed $H$. Let $C$ be a connected component of $H$. Take a subset $S\s V(C)$ containing only regular vertices, and satisfies $|\pt S|\le \al\de$. Then, either $|S|\le 3\al$ or $|S|\ge \de/5$.
\EL
\BP
Since every vertex $v\in S$ is regular, we have $\deg_G(v)\ge\de$, and moreover, since $v$ is not trimmed from $H$, we have $\deg_H(v)\ge\f25\deg_G(v)\ge\f25\de$.
Moreover, since all other vertices in $S$ are regular, at most $|S|-1$ of $v$'s edges can go to another vertex in $S$. Since $\deg_H(v)\ge\f25\de$, this leaves at least $\f25\de-(|S|-1)$ edges to a vertex outside $S$. Altogether, the $|S|$ regular vertices in $S$ are responsible for at least $|S| \cd (\f25\de-(|S|-1))$ many edges in $\pt S$. Therefore,
\begin{align*}
& |S|\cd(\f25\de-(|S|-1)) \le |\pt S| \le \al\de
\\\iff&  -|S|^2 + (\f25\de+1)|S| \le \al\de
\\\iff& |S|^2 - (\f25\de+1)|S| + \al\de \ge 0.
\end{align*}
The solution to $x^2-(\f25\de+1)x+\al\de$ is \[x=\f{(\f25\de+1)\pm\sr{(\f25\de+1)^2-4\al\de}}2 .\]


By \Cref{as:k-de}, $(\f25\de+1)^2-4\al\de\ge0$, so $x$ has real solutions, and $|S|$ must satisfy
\[  |S| \le \f{(\f25\de+1) - \sr{(\f25\de+1)^2 - 4\al\de}}2  \qquad\text{or}\qquad  |S| \ge \f{(\f25\de+1) + \sr{(\f25\de+1)^2 - 4\al\de}}2  .\]
For the second scenario, we clearly have $|S|\ge \de/5$. Now consider the first scenario. We claim that
\[ (\f25\de+1) - \sr{(\f25\de+1)^2 - 4\al\de} \le 6\al,\]
which would imply that $|S|\le 3\al$. This can be seen from
\begin{align*}
& (\f25\de+1)-6\al\le\sr{(\f25\de+1)^2-4\al\de}
\\\iff& (\f25\de+1)^2-12\al(\f25\de+1)+36\al^2\le(\f25\de+1)^2-4\al\de
\\\iff& -0.8\al\de-12\al+36\al^2\le0,
\end{align*}
the last of which follows from \Cref{as:k-de}.
\EP

\BL\label{lem:S-small}
Consider a point in the algorithm's execution, just after it trimmed $H$. Let $C$ be a connected component of $H$ that is a $\g$-expander.
Suppose that $H[C]$ is cut by $\pt_{H[C]}S$ for some $S\s C$ satisfying $|\pt_{H[C]}S|\le \al\de$ and $|S|\le |C|/2$. Then, $S$ has at most $3\al$ regular vertices and no supervertex in $C$.
\EL
\BP
First, suppose for contradiction that $S$ contains a supervertex $s\in C$. Then, since $s$ is active, $\deg_{\ol G}(s)\ge 3\al\de/\g$, and since it's not trimmed, $\deg_H(s)\ge \f25\deg_{\ol G}(s)\ge\f25\cd3\al\de/\g$, so in particular, $\vol_{H[C]}(S)\ge\deg_{H[C]}(s)=\deg_H(s)\ge1.2\al\de/\g$. Therefore, the conductance of the set $S\cap C$ inside $C$ is at most
\[ \f{|\pt_{H[C]}S|}{\vol_{H[C]}(S)} \le \f{\al\de}{1.2\al\de/\g} < \g ,\]
so the set $S$ contradicts the assumption that $H[C]$ is a $\g$-expander.

Now suppose for contradiction that $S$ contains more than $3\al$ regular vertices and no supervertex. By \Cref{lem:S-size-bound}, since $|S|>3\al$, it must be that $|S|\ge\de/5$. Since each vertex in $S$ is regular and is not trimmed, it has degree at least $\f25\de$ in $H$ (\Cref{obs:reg}), so $\vol_{H[C]}(S) \ge |S| \cd \f25\de \ge \f2{25}\de^2$. Again, the conductance of the set $S\cap C$ inside $H[C]$ is at most
\begin{gather} \f{|\pt_{H[C]}S|}{\vol_{H[C]}(S)} \le \f{\al\de}{\f2{25}\de^2} \stackrel{\text{(A\ref{as:k-de})}}< o\lp\f1\logn\rp<\g ,\label{fix2}\end{gather}
so the set $S$ contradicts the assumption that $H[C]$ is a $\g$-expander.
\EP

\BC\label{cor:expander}
Consider a point in the algorithm's execution, just after it trimmed $H$. Let $E^*$ be a $k$-cut of size~$\le \al\de$ in $\ol G$, and let $C$ be a connected component of $H$ that is a $\g$-expander and is cut by $E^*$. Then, all but one component of $E^*$ (in $\ol G$) have at most $3\al$ regular vertices and no supervertices in $C$.
\EC
\BP
Suppose not: there exist two such components that either contain more than $3\al$ regular vertices or one supervertex in $C$. Let $S^*$ be one such component with $|S^*\cap C|\le |C|/2$. Observe that $\pt_{H[C]}(S^*\cap C) \s \pt_{\ol G}S^*$, since every edge going across $S^*\cap C$ in $H[C]$ must also go across $S^*$ in $\ol G$. Therefore,
\[ |\pt_{H[C]}(S^*\cap C)|\le |\pt_{\ol G} S^*| \le \al\de ,\]
so we can apply \Cref{lem:S-small} on the set $S\cap X$, proving the statement.
\EP

\BL\label{lem:no-cut}
Suppose the algorithm is at line~\ref{line:core} of an iteration. Let $A$ be a core of a connected component $C$ that we contract. Let $E^*$ be a nontrivial $k$-cut of size~$\le \al\de$ in $\ol G$. Then, $A$ cannot be cut by $E^*$.
\EL
\BP
Suppose not: $G[A] - OPT$ splits $G[A]$ into more than one component. Let $S^*_0$ be the single component of $OPT$ with $|S^*_0\cap C|>|C|/2$, if it exists, and let $S^*_1,\lds,S^*_r$ be the components of $OPT$ which intersect $A$ and are not $S^*_0$ (by assumption, one must exist). Since the last action of the algorithm before line~\ref{line:core} was trim $H$, by \Cref{cor:expander}, each of $S^*_1,\lds,S^*_r$ has at most $3\al$ regular vertices in $C$ and no supervertex. Let $v\in S^*_1$ be arbitrary; we have $|E[\{v\},S^*_i\cap C]|\le 3\al$ for each $i\ge1$, which means that  $|E[\{v\},(S^*_1\cup\cds\cup S^*_r)\cap C]|\le 3\al k$. Since $v$ was not shaved, we have
\[ \deg_{C}(v) \ge 0.51 \deg_{\ol G}(v) \ge \f12\deg_{\ol G}(v) + 0.01\de \ \stackrel{\mathclap{\text{(A\ref{as:k-de})}}}> \  \f12\deg_{\ol G}(v) + 3\al k\ge\f12\deg_{\ol G}(v)-|E[\{v\},(S^*_1\cup\cds\cup S^*_r)\cap C]| ,\]
so in particular, more than $\f12\deg_{\ol G}(v) $ edges of $v$ go to vertices in $C \setminus (S^*_1 \cup \cds \cup S^*_r)$. This means that $S^*_0$ must exist, and $|E[v,S^*_0]|\ge|E[v,S^*_0\cap C]|>\f12\deg_{\ol G}(v)$. Hence, we also have $ |E[v,S^*_1]| <\f12\deg_{\ol G}(v)$.

 Now consider another $k$-cut formed by moving $v$ from $S^*_1$ to $S^*_0$. This is still a $k$-cut, since the old $k$-cut $E^*$ is nontrivial. Moreover, the value of the new $k$-cut is
\[ |OPT| + |E[v,S^*_1]| - |E[v,S^*_0]| < OPT + \f12\deg_{\ol G}(v) - \f12\deg_{\ol G}(v)= OPT ,\] 
contradicting the choice of $OPT$.
\EP

Finally, \Cref{lem:correctness} easily follows from \Cref{lem:no-cut}, since the only way the lemma can break is if we contract a set of vertices that $OPT$ cuts in line~\ref{line:core}, but this cannot happen by \Cref{lem:no-cut}. This concludes \Cref{lem:correctness}.

\subsection{Quality}\label{sec:KT-q}

In this section, we prove the lemma below (Condition~2 of \Cref{thm:KT}):

\BL[Condition~2 of \Cref{thm:KT}]\label{lem:quality}
At the end of the algorithm, $\ol G$ has $\tO(\al m/\de)$ edges and $\tO(\al m/\de^2)$ vertices.
\EL

We first introduce two lemmas, one directly from \cite{KT}, and one we reprove:

\BL[Lemma 17 of \cite{KT}]\label{lem:17}
There are $\Om(\de^2)$ edges from $G$ contracted in each supervertex of $\ol G$.
\EL

\BL[Lemma 18 of \cite{KT}, reproven]\label{lem:num-edges}
The total number of edges leaving passive supervertices is $O(\al\log n \cd m/\de)$.
\EL
\BP
By \Cref{lem:17}, every passive supervertex has $\Om(\de^2)$ edges contracted to it, and since there are $m$ edges total, there are $O(m/\de^2)$ passive supervertices. By definition, a passive supervertex has degree~$\le 3\al\de/\g$, which is at most $O(m/\de^2)\cd3\al\de/\g=O(\al\log m \cd m/\de)$.
\EP
We now prove \Cref{lem:quality}.
Since \ref{KT} terminates when a constant fraction of the edges of $\ol G$ are incident to passive supervertices, we conclude that when the algorithm terminates, $\ol G$ has $O(\al\log m\cd m/\de)$ edges. We now focus on the vertex bound of \Cref{lem:quality}. Since each regular vertex in $\ol G$ must have degree $\ge \de$, there are at most $O(\al\log m \cd m/\de^2)$ many of them. From the proof of \Cref{lem:num-edges}, there are $O(m/\de^2)$ supervertices in $\ol G$, so altogether, $\ol G$ has $O(\al\log m\cd m/\de^2)$ vertices. This concludes the proof of \Cref{lem:quality}.


\subsection{Termination}\label{sec:KT-t}

In \cite{KT}, since they aimed at a near-linear time algorithm, they needed the graph $\ol G$ to shrink by a constant factor per iteration. Here, all we need is \emph{some} progress in $\ol G$, i.e., one single contraction, so that the algorithm does not loop indefinitely. Nevertheless, we will still prove that the number of edges of $\ol G$ decreases by a constant factor:
\BL[Lemma~A.5 of \cite{flow}] \label{lem:termination}
In each except for the last iteration of the outer loop, i.e., the repeat loop, the number of edges in the graph $\ol G$ is decreased by a factor of at least $7/10$.
\EL

The following lemma, which relates the total number of edges cut during an iteration to the low-conductance cuts in line~\ref{line:cut}, is mostly unchanged.
\BL[Lemma~A.4 of \cite{flow}]\label{lem:A4}
If the total number of edges cut in line~\ref{line:cut} during an iteration of the outer loop is $c$, then the total number of edges lost from all clusters due to trimming, shaving, and scrapping during this iteration is $6c$.
\EL

The only part of its proof that is different is the scrapping part, since we defined loose vertices slightly differently from \cite{KT}. We prove our version of this part below, which is conveniently captured as Lemma~A.3 in \cite{flow}. (The use of $k$ in Lemma~A.3 clashes with our notion of $k$ (in $k$-cut), so we changed it to $k'$ instead.)

\BL[Lemma~A.3 of \cite{flow}, reproven]
If a component $C$ has $k'$ edges leaving it in $\ol G$ and the core of $C$ is scrapped, then $\vol_{\ol G}(C)\le4k'$.
\EL
\BP
Call an edge with at least one endpoint in $C$ \emph{internal to the core} if both of its endpoints are inside the core.
We first prove that there are always at most $3k'$ edges in $C$ that are not internal to the core. There are two types of edges incident to $C$ and not internal to the core: the edges incident to loose  vertices in $C$ and the edges in $E[C,\ol G-C]$. For the first type, since every loose vertex $v\in C$ has $|E[v,V(\ol G)\setminus C]| \ge 0.49\deg_{\ol G}(v)$, the number of edges of the first type, denoted $n_1$, is at most
\[ \sum_{\text{loose }v}\deg_{\ol G}(v) \le \f1{0.49}\sum_{\text{loose }v}|E[v,V(\ol G)\setminus C]| .\]
But the number of second type edges, denoted $n_2$, is exactly $k'-\sum_{\text{loose }v}|E[v,V(\ol G)\setminus C]|$, so altogether, there are
\[n_1+n_2\le \f1{0.49}\sum_{\text{loose }v}|E[v,V(\ol G)\setminus C]|+ \lp k-\sum_{\text{loose }v}|E[v,V(\ol G)\setminus C]| \rp \le \f1{0.49}k' \le2.1k'<3k' \]
many edges not internal to the core.

Now suppose the core $A$ is scraped. By definition, this means that $\vol_{H[C]}(A)\le\vol_{\ol G}(C)/4$. Now observe that $\vol_{\ol G}(C)$ is exactly $\vol_{H[C]}(A) + n_1 + n_2$, so
\[ \vol_{\ol G}(C)=\vol_{H[C]}(A)+n_1+n_2\le\vol_{\ol G}(C)/4+3k \implies \vol_{\ol G}(C)\le4k' .\]
\EP

We now bound the number of edges cut by low-conductance cuts (line~\ref{line:cut}). Instead of using the more complicated procedures in \cite{flow,KT}, we resort to simply iteratively computing low-conductance cuts. As for why the low-conductance cuts cut a small number of edges in total, the reasoning is the same as the one in \cite{KT}, and we sketch it here for convenience.

\BL\label{lem:cond-cuts}
For an appropriate $\g$ in the definition of $\g$, the total number of edges cut in line~\ref{line:cut} during an iteration of the outer loop is at most $0.04|E(\ol G)|$.
\EL
\BP
We set up a charging scheme as follows: every time we compute a low-conductance cut $E'$ in a connected component $C$ of $H$, we charge a total cost of $|E'|$ uniformly to the edges of the smaller side $S$ of $C$ (the side with smaller $\vol_{\ol G}(S)$). Since $E'$ has conductance~$\le\g$, we have
\[ \f{|E'|}{\vol_{\ol G}(S)} = \f{|E'|}{2|E[S]|+|E'|}\le\g \implies |E'|\le\f{2\g}{1-\g}|E[S]| \le 4\g|E[S]|,\]
so every edge in $E[S]$ is charged at most $4\g$. Now observe that every time an edge is charged, since it belongs to the smaller side of the cut, the size of the component containing this edge halves, so an edge is charged at most $\log m$ times. In total, an edge is charged a cost of at most $4\g\log m$. Therefore, at most $4\g |E(\ol G)| \log m=0.04|E(\ol G)|$ cost was charged in total, and this also upper bounds the total edges cut.
\EP

Therefore, by \Cref{lem:A4,lem:cond-cuts}, the algorithm cuts at most $7 \cd 0.04|E(\ol G)|$ edges in $\ol G$ on a given iteration. Since it contracts the rest, we have proven \Cref{lem:termination}.

\subsection{Polynomial-time Algorithm}

Next, we modify the algorithm \ref{KT} to take an approximate conductance cut algorithm instead, making it run in polynomial time. We use the $O(\sr{\logn})$-approximation algorithm of Arora, Rao, and Vazirani  below:

\BT[\cite{arora2009expander}]
There exists a universal constant $C>0$ and a polynomial-time $C\sr{\log m}$-approximation algorithm for minimum conductance cut.
\ET
\begin{algorithm}[H]
\mylabel{KTP}{\texttt{KT-Sparsification-Polytime}}
\caption{\ref{KTP}($G=(V,E),\al$)}
\begin{algorithmic}[1]
\State $\ol G\gets G$; $G$ has min degree $\de$
\While{$<\f1{20}$ fraction of the edges in $\ol G$ are incident to passive supervertices}\label{line:2}
  \State $H\gets \ol G$
  \State Remove passive supervertices from $H$ and trim $H$
  \While{there exists a connected component $C$ in $H$ such that $H[C]$ is not a $\g$-expander}
    \State Compute a $\textcolor{blue}{C\sr{\log m}} \cd\g$-conductance cut in $H[C]$, remove these edges from $H$, and trim $H$\label{line:cut}
  \EndWhile
  \State Take each connected component of $H$ and contract its core (if nonempty) to a supervertex in $\ol G$\label{line:core}
\EndWhile
\State\Return$\ol G$
\end{algorithmic}\label{alg:KT}
\end{algorithm}

The entire analysis goes through without change, except for the following differences:

\BE
\im We re-define the parameter $\g:=\f1{100C\log^{1.5}m}$ throughout the entire algorithm and analysis.
\im \Cref{as:k-de} is replaced by the assumption that
\begin{gather} \de>\om(\max\{\al\log^{1.5}n,\al k\}) \label{fix3}\end{gather}
instead.
\im In the proof of \Cref{lem:S-small}, \Cref{fix2} is replaced with
\[ \f{|\pt_{H[C]}S|}{\vol_{H[C]}(S)} \le \f{\al\de}{\f2{25}\de^2} \stackrel{(\ref{fix3})}< o\lp\f1{\log^{1.5}n}\rp<\g  ,\]
and the rest of the proof of \Cref{lem:S-small} is identical.
\im In the proof of \Cref{lem:cond-cuts}, every instance of $\g$ is replaced by $C\sr{\log m}\cd\g$ (with the new value of $\g)$. Since $C\sr{\log m}\cd\g=\f1{100\log m}$, which is exactly the old value of $\g$, the rest of the proof of \Cref{lem:cond-cuts} is unchanged.
\EE



\bibliography{ref}

\begin{thebibliography}{10}

\bibitem{abboud2015if}
Amir Abboud, Arturs Backurs, and Virginia~Vassilevska Williams.
\newblock If the current clique algorithms are optimal, so is {V}aliant's
  parser.
\newblock In {\em Foundations of Computer Science (FOCS), 2015 IEEE 56th Annual
  Symposium on}, pages 98--117. IEEE, 2015.

\bibitem{alon1995color}
Noga Alon, Raphael Yuster, and Uri Zwick.
\newblock Color-coding.
\newblock {\em J. ACM}, 42(4):844--856, 1995.

\bibitem{arora2009expander}
Sanjeev Arora, Satish Rao, and Umesh Vazirani.
\newblock Expander flows, geometric embeddings and graph partitioning.
\newblock {\em Journal of the ACM (JACM)}, 56(2):5, 2009.

\bibitem{chekuri2018lp}
Chandra Chekuri, Kent Quanrud, and Chao Xu.
\newblock Lp relaxation and tree packing for minimum $ k $-cuts.
\newblock {\em arXiv preprint arXiv:1808.05765}, 2018.

\bibitem{Chitnis}
Rajesh Chitnis, Marek Cygan, MohammadTaghi Hajiaghayi, Marcin Pilipczuk, and
  Micha{\l} Pilipczuk.
\newblock Designing {FPT} algorithms for cut problems using randomized
  contractions.
\newblock {\em SIAM J. Comput.}, 45(4):1171--1229, 2016.
\newblock URL: \url{http://dx.doi.org/10.1137/15M1032077}, \href
  {http://dx.doi.org/10.1137/15M1032077} {\path{doi:10.1137/15M1032077}}.

\bibitem{FPT-book}
Marek Cygan, Fedor~V. Fomin, {\L}ukasz Kowalik, Daniel Lokshtanov, D\'aniel
  Marx, Marcin Pilipczuk, Micha{\l} Pilipczuk, and Saket Saurabh.
\newblock {\em Parameterized algorithms}.
\newblock Springer, Cham, 2015.
\newblock URL: \url{http://dx.doi.org/10.1007/978-3-319-21275-3}, \href
  {http://dx.doi.org/10.1007/978-3-319-21275-3}
  {\path{doi:10.1007/978-3-319-21275-3}}.

\bibitem{DEFPR03}
Rodney~G. Downey, Vladimir Estivill-Castro, Michael Fellows, Elena Prieto, and
  Frances~A. Rosamund.
\newblock Cutting up is hard to do: The parameterised complexity of $k$-cut and
  related problems.
\newblock {\em Electronic Notes in Theoretical Computer Science}, 78:209--222,
  2003.

\bibitem{GH94}
Olivier Goldschmidt and Dorit~S. Hochbaum.
\newblock A polynomial algorithm for the {$k$}-cut problem for fixed {$k$}.
\newblock {\em Math. Oper. Res.}, 19(1):24--37, 1994.
\newblock URL: \url{http://dx.doi.org/10.1287/moor.19.1.24}, \href
  {http://dx.doi.org/10.1287/moor.19.1.24} {\path{doi:10.1287/moor.19.1.24}}.

\bibitem{GLL18b}
Anupam Gupta, Euiwoong Lee, and Jason Li.
\newblock Faster exact and approximate algorithms for $k$-cut.
\newblock In {\em Foundations of Computer Science (FOCS), 2018 IEEE 59th Annual
  Symposium on}, 2018.

\bibitem{GLL19}
Anupam Gupta, Euiwoong Lee, and Jason Li.
\newblock The number of minimum $k$-cuts: Improving the karger-stein bound.
\newblock In {\em STOC 2019, to appear}, 2019.

\bibitem{flow}
Monika Henzinger, Satish Rao, and Di~Wang.
\newblock Local flow partitioning for faster edge connectivity.
\newblock In {\em Proceedings of the Twenty-Eighth Annual ACM-SIAM Symposium on
  Discrete Algorithms}, pages 1919--1938. Society for Industrial and Applied
  Mathematics, 2017.

\bibitem{KYN06}
Yoko Kamidoi, Noriyoshi Yoshida, and Hiroshi Nagamochi.
\newblock A deterministic algorithm for finding all minimum {$k$}-way cuts.
\newblock {\em SIAM J. Comput.}, 36(5):1329--1341, 2006/07.
\newblock URL: \url{http://dx.doi.org/10.1137/050631616}, \href
  {http://dx.doi.org/10.1137/050631616} {\path{doi:10.1137/050631616}}.

\bibitem{Kapoor96}
Sanjiv Kapoor.
\newblock On minimum {$3$}-cuts and approximating {$k$}-cuts using cut trees.
\newblock In {\em Integer programming and combinatorial optimization
  ({V}ancouver, {BC}, 1996)}, volume 1084 of {\em Lecture Notes in Comput.
  Sci.}, pages 132--146. Springer, Berlin, 1996.
\newblock URL: \url{http://dx.doi.org/10.1007/3-540-61310-2_11}, \href
  {http://dx.doi.org/10.1007/3-540-61310-2_11}
  {\path{doi:10.1007/3-540-61310-2_11}}.

\bibitem{KS96}
David~R. Karger and Clifford Stein.
\newblock A new approach to the minimum cut problem.
\newblock {\em Journal of the ACM (JACM)}, 43(4):601--640, 1996.

\bibitem{KT11}
Ken-ichi Kawarabayashi and Mikkel Thorup.
\newblock The minimum $k$-way cut of bounded size is fixed-parameter tractable.
\newblock In {\em Foundations of Computer Science (FOCS), 2011 IEEE 52nd Annual
  Symposium on}, pages 160--169. IEEE, 2011.

\bibitem{KT}
Ken-ichi Kawarabayashi and Mikkel Thorup.
\newblock Deterministic edge connectivity in near-linear time.
\newblock {\em Journal of the ACM (JACM)}, 66(1):4, 2018.

\bibitem{le2014powers}
Fran{\c{c}}ois Le~Gall.
\newblock Powers of tensors and fast matrix multiplication.
\newblock In {\em Proceedings of the 39th international symposium on symbolic
  and algebraic computation}, pages 296--303. ACM, 2014.

\bibitem{Manurangsi17}
Pasin Manurangsi.
\newblock {Inapproximability of Maximum Edge Biclique, Maximum Balanced
  Biclique and Minimum $k$-Cut from the Small Set Expansion Hypothesis}.
\newblock In {\em 44th International Colloquium on Automata, Languages, and
  Programming (ICALP 2017)}, volume~80 of {\em Leibniz International
  Proceedings in Informatics (LIPIcs)}, pages 79:1--79:14, 2017.
\newblock URL: \url{http://drops.dagstuhl.de/opus/volltexte/2017/7500}, \href
  {http://dx.doi.org/10.4230/LIPIcs.ICALP.2017.79}
  {\path{doi:10.4230/LIPIcs.ICALP.2017.79}}.

\bibitem{Marx07}
D{\'a}niel Marx.
\newblock Parameterized complexity and approximation algorithms.
\newblock {\em The Computer Journal}, 51(1):60--78, 2007.

\bibitem{NI92}
Hiroshi Nagamochi and Toshihide Ibaraki.
\newblock Computing edge-connectivity in multigraphs and capacitated graphs.
\newblock {\em SIAM J. Discrete Math.}, 5(1):54--66, 1992.
\newblock URL: \url{http://dx.doi.org/10.1137/0405004}, \href
  {http://dx.doi.org/10.1137/0405004} {\path{doi:10.1137/0405004}}.

\bibitem{NR01}
Joseph Naor and Yuval Rabani.
\newblock Tree packing and approximating {$k$}-cuts.
\newblock In {\em Proceedings of the {T}welfth {A}nnual {ACM}-{SIAM}
  {S}ymposium on {D}iscrete {A}lgorithms ({W}ashington, {DC}, 2001)}, pages
  26--27. SIAM, Philadelphia, PA, 2001.

\bibitem{RS02}
R.~Ravi and Amitabh Sinha.
\newblock Approximating {$k$}-cuts using network strength as a {L}agrangean
  relaxation.
\newblock {\em European J. Oper. Res.}, 186(1):77--90, 2008.
\newblock URL: \url{http://dx.doi.org/10.1016/j.ejor.2007.01.040}, \href
  {http://dx.doi.org/10.1016/j.ejor.2007.01.040}
  {\path{doi:10.1016/j.ejor.2007.01.040}}.

\bibitem{SV95}
Huzur Saran and Vijay~V. Vazirani.
\newblock Finding $k$-cuts within twice the optimal.
\newblock {\em SIAM Journal on Computing}, 24(1):101--108, 1995.

\bibitem{Thorup08}
Mikkel Thorup.
\newblock Minimum $k$-way cuts via deterministic greedy tree packing.
\newblock In {\em Proceedings of the fortieth annual ACM symposium on Theory of
  computing}, pages 159--166. ACM, 2008.

\bibitem{williams2012multiplying}
Virginia~Vassilevska Williams.
\newblock Multiplying matrices faster than {Coppersmith}--{Winograd}.
\newblock In {\em Proceedings of the forty-fourth annual ACM symposium on
  Theory of computing}, pages 887--898. ACM, 2012.

\bibitem{williams2010subcubic}
Virginia~Vassilevska Williams and Ryan Williams.
\newblock Subcubic equivalences between path, matrix and triangle problems.
\newblock In {\em Foundations of Computer Science (FOCS), 2010 51st Annual IEEE
  Symposium on}, pages 645--654. IEEE, 2010.

\bibitem{XCY11}
Mingyu Xiao, Leizhen Cai, and Andrew Chi-Chih Yao.
\newblock Tight approximation ratio of a general greedy splitting algorithm for
  the minimum $k$-way cut problem.
\newblock {\em Algorithmica}, 59(4):510--520, 2011.

\bibitem{ZNI01}
Liang Zhao, Hiroshi Nagamochi, and Toshihide Ibaraki.
\newblock Approximating the minimum {$k$}-way cut in a graph via minimum 3-way
  cuts.
\newblock {\em J. Comb. Optim.}, 5(4):397--410, 2001.
\newblock URL: \url{http://dx.doi.org/10.1023/A:1011620607786}, \href
  {http://dx.doi.org/10.1023/A:1011620607786}
  {\path{doi:10.1023/A:1011620607786}}.

\end{thebibliography}

\section{Proof of ~\Cref{thm:lower}}

Given a $(k-1)$-clique graph instance $G=(V,E)$, construct the following graph $H$: Let $W$ be a clique of size $k^2n$. Take the union of $G$ and $W$, and then for each vertex $v\in V$, add $n-\deg(v)$ edges to arbitrary vertices in $W$. This is the graph $H$.  Note that a $k$-cut of size~$\le (k-1)n$ can be formed by isolating any $(k-1)$ vertices in $G$. We now claim that $G$ has a $(k-1)$-clique iff $H$ has minimum $k$-cut $(k-1)n-\bn{k-1}2$. 

Fix a minimum $k$-cut $S^*_1,\lds, S^*_k$ in $H$. First, if $W$ is not entirely contained in one component, then the cut has size~$\ge k^2n-1$ already, so we can assume that this does not happen. Let us assume that $W$ is contained in $S^*_k$. Next, if $|S^*_1|+|S^*_2|+\cds+|S^*_{k-1}|>k-1$, then this $k$-cut will have size~$\ge kn-\bn{k-1}2 > (k-1)n$ (we can assume $n\gg k$), so this cannot happen either. Therefore, $S^*_i$ is a singleton vertex $v^*_i\in V$ for all $i\in[k-1]$. It follows that the minimum $k$-cut has cost exactly $(k-1)n-|E[\{v^*_1\},\lds,\{v^*_{k-1}\}]|$. This is exactly $(k-1)n-\bn{k-1}2$ iff $G$ has a $(k-1)$-clique.

\end{document}